%% file: main.tex
\documentclass{article}

\usepackage[final]{neurips_2025}
\usepackage[utf8]{inputenc} % allow utf-8 input
\usepackage[T1]{fontenc}    % use 8-bit T1 fonts
\usepackage{hyperref}       % hyperlinks
\usepackage{url}            % simple URL typesetting
\usepackage{booktabs}       % professional-quality tables
\usepackage{amsfonts}       % blackboard math symbols
\usepackage{nicefrac}       % compact symbols for 1/2, etc.
\usepackage{microtype}      % microtypography
\usepackage{graphicx}

\usepackage{amsthm}

\newtheorem{theorem}{Theorem}

\usepackage{xspace}

\newcommand{\xm}[1]{\textcolor{black}{#1}}
\newcommand{\name}{\texttt{ObCLIP}\xspace}

\usepackage{pifont}
\usepackage{caption}
\usepackage[table,xcdraw]{xcolor}
\usepackage{amsmath,amsfonts,amssymb,stmaryrd,xspace,enumitem}

\usepackage{graphicx}
\usepackage{caption}
\usepackage{float}  % 为了使用 [H] 强制位置
\usepackage{multicol}
\usepackage{multirow}
\usepackage{amsmath}
\usepackage{subcaption}
\usepackage{wrapfig}
\usepackage{listings}
\usepackage[linesnumbered,ruled,vlined]{algorithm2e} 

\usepackage{pifont}
\usepackage{tikz}
\newcommand*\emptycirc[1][0.8ex]{\tikz\draw (0,0) circle (#1);} 
\newcommand*\halfcirc[1][0.8ex]{%
	\begin{tikzpicture}
	\draw[fill] (0,0)-- (90:#1) arc (90:270:#1) -- cycle ;
	\draw (0,0) circle (#1);
	\end{tikzpicture}}
\newcommand*\fullcirc[1][0.8ex]{\tikz\fill (0,0) circle (#1);} 

\begin{document}

\title{\name: \underline{Ob}livious \underline{CL}oud-Device Hybrid \underline{I}mage Generation with \underline{P}rivacy Preservation}

\author{
    Haoqi Wu\textsuperscript{\rm 1,\thanks{Email at haoqi.1997@tiktok.com}}, 
    Wei Dai\textsuperscript{\rm 1},
    Ming Xu\textsuperscript{\rm 2}, 
    Li Wang\textsuperscript{\rm 1},
    Qiang Yan\textsuperscript{\rm 1} \\
    \textsuperscript{\rm 1}TikTok Inc., \textsuperscript{\rm 2}National University of Singapore \\
    }

\maketitle

\begin{abstract} 

Diffusion Models have gained significant popularity due to their remarkable capabilities in image generation, albeit at the cost of intensive computation requirement. Meanwhile, despite their widespread deployment in inference services such as Midjourney, concerns about the potential leakage of sensitive information in uploaded user prompts have arisen.
Existing solutions either lack rigorous privacy guarantees or fail to strike an effective balance between utility and efficiency. 
To bridge this gap, we propose \name, a plug-and-play safeguard that enables \textit{oblivious} cloud-device hybrid generation.
By \textit{oblivious}, each input prompt is transformed into a set of semantically similar candidate prompts that differ only in sensitive attributes (e.g., gender, ethnicity).
The cloud server processes all candidate prompts without knowing which one is the real one, thus preventing any prompt leakage.
To mitigate server cost, only a small portion of denoising steps is performed upon the large cloud model. The intermediate latents are then sent back to the client, which selects the targeted latent and completes the remaining denoising using a small device model.
Additionally, we analyze and incorporate several cache-based accelerations that leverage temporal and batch redundancy, effectively reducing computation cost with minimal utility degradation.
Extensive experiments across multiple datasets demonstrate that \name provides rigorous privacy and comparable utility to cloud models with slightly increased server cost.
\end{abstract}

\input{tex/1-introduction}
\input{tex/2-related}

\input{tex/3-preliminary}
\input{tex/4-design}
\input{tex/5-experiments}
\input{tex/6-conclusion}

\bibliographystyle{plain}
\bibliography{sample-base}

\newpage
\input{tex/7-checklist}

\newpage
\appendix
\input{tex/appendix}

\end{document}

%% file: tex/1-introduction.tex
\section{Introduction}\label{sec:intro}
\begin{figure}[h]
    \centering
    \setlength{\abovecaptionskip}{1pt}
    \setlength{\belowcaptionskip}{0pt}
    \includegraphics[width=0.75\linewidth]{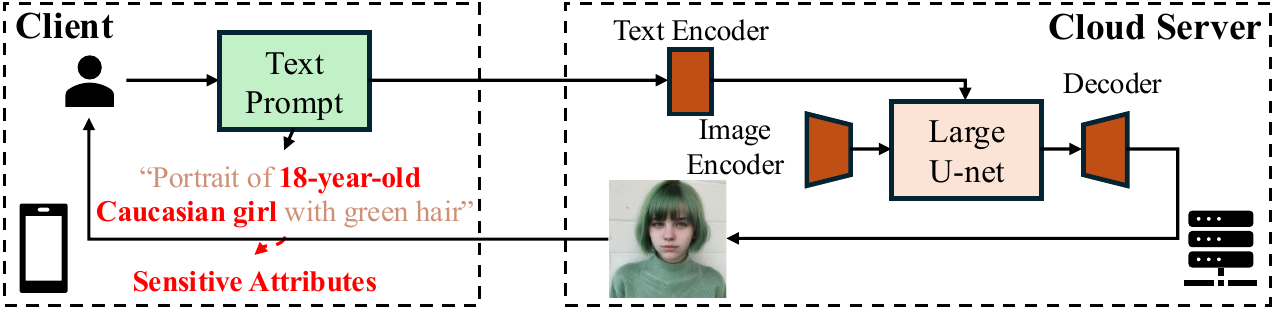}
    \caption{Illustration of existing server-only text-to-image generation services.}
    \label{fig:scenario-illustration}
\end{figure}

\xm{Stable diffusion models~\cite{ldm-22, sdxl-24} have emerged as a de-facto standard technique in text-to-image (T2I) generation due to their superior capability to generate high-quality images. This drives the widespread application of T2I in inference services hosted by cloud servers.}
As depicted in Figure~\ref{fig:scenario-illustration}, the client uploads text prompts to the cloud, which generates images and sends them back to the client. 
This paradigm is widely-adopted since the generation typically requires huge computation cost, which is unaffordable for clients, especially for devices with limited computation power. 

However, despite its growing popularity, there remain two essential problems: 
\ding{172} \textit{\textbf{Prompt privacy}} remains a critical concern~\cite{mmdt-25, benchmark-sensitive-llm-outputs-25}: In image generation services like Midjourney~\cite{midjourney2025} and DALL·E~\cite{dalle2025}, the server deploys their models on the cloud and provides APIs that take client prompts as inputs, which might contain sensitive attributes like \textit{gender, ethnicity}, etc. 
\ding{173} \textbf{\textit{Server cost}} increases drastically. According to the scaling law~\cite{scaling-law, scaling-law-diffusion-24}, the superior model capacity comes at the expense of larger model size, leading to considerable hardware requirement and computation cost on the cloud servers.

Existing solutions typically only address one of the aforementioned issues. Cryptographic-based approaches~\cite{mpcvit-23, he-diffusion-24} offer provable security to the sensitive prompts. However, the huge computation overhead and efficiency decay hinders their application in real world scenarios. Meanwhile, some prior works~\cite{santext-21, casper-text-prompt-privacy} provide a client-side input filter that detects and perturbs sensitive information before sent to the server. However, such kind of random perturbation incurs inevitable semantic and utility loss, leading to mismatch between the user intent and generation output. 
Another line of works employ on-device models~\cite{snapfusion-23, mobilediffusion-24} to allow efficient image generation, avoiding data transmission to the server.
However, despite significant improvement that has been made, the image quality inevitably decreases, failing to meet the users' needs.
Recently, Hybrid SD~\cite{hybridsd-24} incorporates a hybrid image generation pipeline to lower the server-side computation cost, which however fails to preserve the prompt privacy, the embedding of which is directly sent to the server. Such kind of information is vulnerable to extraction attacks~\cite{embedding-attack-sp20, text-embedding-inversion-23}.
Therefore, here arises the question: 
\begin{center}
    \textit{Can we perform privacy-preserving image generation with better image quality and lower server cost?}
\end{center}

As an attempt to answer this question, we propose \name, an oblivious cloud-device hybrid image generation scheme that provides rigorous privacy and comparable utility to large cloud models with slightly increased server cost.
Specifically, our contributions are summarized as follows: 
\begin{itemize}[itemsep=0em, topsep=0em, leftmargin=2em]
    \item \textbf{Oblivious Cloud-Device Hybrid Generation Scheme}. \name consists of two main components \xm{to address the aforementioned challenges}:
    1) \textit{Oblivious transformation}:  each input prompt is transformed into a set of semantically similar candidate prompts that differ only in sensitive attributes (e.g., gender ,age, ethnicity).
    The cloud server processes all candidate prompts without knowing which one is the real one, thus preventing any prompt leakage.
    2) \textit{Local extraction}: the client selects the targeted output corresponding to the real prompt.
    One straightforward drawback of vanilla oblivious generation is heavy server cost. Therefore, we devise a hybrid generation pipeline, where only partial denoising steps are performed by the server. Besides, we analyze and incorporate several cache-based acceleration methods, leveraging both temporal and batch redundancy, to further reduce server cost with minimal utility degradation. 
    
    \item \textbf{Temporal- and Batch- Redundancy based Acceleration}. We analyze and incorporate several cache-based acceleration techniques to exploit temporal redundancy in server-side generation. Additionally, inspired by batch-level redundancy, we propose reusing attention maps across the batch of candidate prompts. Together, these two acceleration strategies effectively reduce computation costs with minimal utility degradation.
    
    \item \textbf{Empirical Evaluations.} We conduct extensive text-to-image generation experiments on several stable diffusion models across three datasets.
    The experiments confirm that \name provides rigorous privacy and comparable utility to large cloud models with slightly increased server computation costs, which is about $4.4\sim 7.6\times$ lower than vanilla oblivious generation baseline and orders of magnitude lower than cryptographic approach.
\end{itemize}

%% file: tex/2-related.tex
\section{Related Work}\label{sec:related}
Existing works typically employ various privacy enhancing technologies.
We provide a comprehensive comparison of related works in Table~\ref{tab:comparison}, focusing on application domain and trade-off among prompt privacy, server cost and image utility.

\begin{table}[ht]
\setlength{\abovecaptionskip}{0pt}
\setlength{\belowcaptionskip}{0pt}
\renewcommand\tabcolsep{5.3pt} 
    \centering
    \caption{Comparison of related work. $\fullcirc$, $\halfcirc$ and $\emptycirc$ refer to high-, medium- and low-performance.}
    \scalebox{0.8}{
    \begin{tabular}{@{}ccc|cccc@{}}
\toprule
\multicolumn{3}{c|}{\textbf{Method}}                                                                                                                          & \textbf{Domain} & \textbf{Privacy} & \textbf{Server Cost} & \textbf{Utility} \\ \midrule
\multicolumn{1}{c|}{\multirow{3}{*}{Non-private}} & \multicolumn{1}{c|}{\multirow{2}{*}{Standalone}} & Server-Only$~\cite{sdxl-24}$                           & Text-to-Image   & $\emptycirc$     & $\emptycirc$         & $\fullcirc$      \\
\multicolumn{1}{c|}{}                             & \multicolumn{1}{c|}{}                            & Client-Only$~\cite{snapfusion-23, mobilediffusion-24}$ & Text-to-Image   & $\fullcirc$      & $\fullcirc$          & $\emptycirc$     \\ \cmidrule(l){2-7} 
\multicolumn{1}{c|}{}                             & \multicolumn{1}{c|}{Hybrid}                      & Hybrid SD$~\cite{hybridsd-24}$                         & Text-to-Image   & $\emptycirc$     & $\halfcirc$          & $\halfcirc$      \\ \midrule
\multicolumn{1}{c|}{\multirow{5}{*}{Private}}     & \multicolumn{1}{c|}{\multirow{2}{*}{MPC}}        & MPCViT$~\cite{mpcvit-23}$                              & Text-to-Image   & $\fullcirc$      & $\emptycirc$         & $\halfcirc$      \\
\multicolumn{1}{c|}{}                             & \multicolumn{1}{c|}{}                            & HE-Diffusion$~\cite{he-diffusion-24}$                  & Text-to-Image   & $\fullcirc$      & $\emptycirc$         & $\halfcirc$      \\ \cmidrule(l){2-7} 
\multicolumn{1}{c|}{}                             & \multicolumn{1}{c|}{\multirow{2}{*}{DP}}         & SANTEXT$~\cite{santext-21}$                            & Text Generation & $\halfcirc$      & $\fullcirc$          & $\emptycirc$     \\
\multicolumn{1}{c|}{}                             & \multicolumn{1}{c|}{}                            & CAPE$~\cite{cape-icml25}$                              & Text Generation & $\halfcirc$      & $\fullcirc$          & $\emptycirc$     \\ \cmidrule(l){2-7} 
\multicolumn{1}{c|}{}                             & \multicolumn{1}{c|}{Ours}                        & $\name$                                                & Text-to-Image   & $\fullcirc$      & \textbf{$\halfcirc$} & $\halfcirc$      \\ \bottomrule
\end{tabular}
}
    \label{tab:comparison}
\end{table}

Cryptographic methods like secure multi-party computation (MPC) that support computation over encrypted data and models are widely used to enable secure machine learning inference~\cite{spu, puma-2023, ditto-24, mpcvit-23}.
MPCViT~\cite{mpcvit-23} employed MPC and proposed a search algorithm for MPC-friendly neural architecture.
HE-Diffusion~\cite{he-diffusion-24} leveraged homomorphic encryption (HE) to perform partial image encryption and protected the diffusion process.
However, these works impose significant overhead compared to plaintext baselines, limiting their practicality for real-world deployment.
Other works employ lightweight privacy techniques like differential privacy (DP)~\cite{dp-06} to perturb prompts by adding random noises in text generation. SANTEXT~\cite{santext-21} designed a Exponential mechanism based word-level perturbation mechanism to hide sensitive attributes within a prompt. CAPE~\cite{cape-icml25} further proposed an optimized perturbation mechanism by incorporating contextual information that achieves a better trade-off between privacy and utility. However, privacy comes at the cost of semantics distortion and utility degradation. Besides, the scalability to text-to-image generation domain is uncertain.
Another line of works manage to offload the generation to user devices, avoiding the transmission of prompts to server. SnapFusion~\cite{snapfusion-23} introduced efficient network architecture and improved step distillation process to enable generation within 2 seconds. MobileDiffusion~\cite{mobilediffusion-24} further studied one-step sampling technique, which decreased the runtime to less than 1 second. However, its scalability to larger models like SDXL~\cite{sdxl-24} is not yet explored.

Recently, a new paradigm of cloud-device hybrid generation scheme is proposed to lower the server-side computation cost. Hybrid SD~\cite{hybridsd-24} proposed an cloud-device collaborative stable diffusion pipeline. By offloading a great portion of denoising steps to client devices, the server-side costs can be optimized. 
However, it only reduces server costs, failing to effectively protect the privacy of user prompts, the embedding of which is directly sent to the server. Such kind of information is vulnerable to extraction attacks~\cite{collin2025universal-zs-embedding-inversion, text-embedding-inversion-23, embedding-attack-sp20}, which could reconstruct the original prompt or infer partial sensitive attributes.
To make matters worse, even if embedding inversion fails, the server can still perform full image generation based on the received embeddings, inevitably revealing sensitive visual patterns.
Built upon such paradigm, we manage to preserve the privacy in an oblivious way, providing rigorous privacy. To hedge against the additional server-side cost introduced by our scheme, we incorporate several acceleration methods based on temporal and batch redundancy.

%% file: tex/3-preliminary.tex
\section{Preliminary}\label{sec:preliminary}
\noindent\textbf{Diffusion Model.}\label{sec:diffusion-model}
Diffusion models~\cite{ddpm-20} have attracted significant attentions due to their ability to generate high‐quality images.
Its forward process adds Gaussian noise to the data over a fixed number of timesteps as $q(x_t|x_{t-1}) = \mathcal{N}(x_t; \sqrt{1-\beta_t}\cdot x_{t-1}, \beta_t\mathbf{I})$, where $x_0$ refers to the original image, $t \in [1, ..., T]$ denotes the total diffusion timestep, $\{x_1,...,x_{T}\}$ denote the sequence of noisy latents, $\beta_t \in (0, 1)$ determines the amount of noise added at each timestep, $\mathbf{I}$ is the identity matrix and $\mathcal{N}(x;\mu, \sigma)$ denotes the Gaussian distribution with mean $\mu$ and covariance $\sigma$.
During image generation, the reverse (denoising) process aims to recover $x_{t-1}$ from $x_t$ using a neural network as the noise predictor (typically, U-net~\cite{unet-15}) $\epsilon_\theta(x_t, t)$ that predicts the noise in each timestep. Concretely,
\begin{equation}
    x_{t-1} = \frac{1}{\sqrt{\alpha_t}} \left( x_t - \frac{\beta_t}{\sqrt{1 - \bar{\alpha}_t}} \epsilon_\theta(x_t, t) \right) + \sqrt{\beta_t} z
\end{equation}
where $\alpha_t = 1 - \beta_t$ and $\bar{\alpha}_t = \prod_{i=1}^{T}\alpha_i$.
This process is iteratively applied starting from a random sample drawn from the noise prior and finally yields a denoised sample.
For stable diffusion models~\cite{ldm-22}, the core idea is to perform the diffusion process in a lower-dimensional latent space—obtained via a pretrained variational autoencoder (VAE)—rather than directly in pixel space. This approach significantly reduces computation overhead during denoising process.
Recently, the MMDiT architecture~\cite{mmdit, seeddream3}, which jointly processes image and text tokens to model cross-modal relationships, has also been gaining increasing attention.

\noindent\textbf{Training-free Diffusion Acceleration.}
To mitigate the expensive computation cost of diffusion generation, prior works either resort to training-based model distillation~\cite{step-distillation-22, lcm-23} and model compression~\cite{model-pruning-23, bk-sdm-24} approaches, or training-free caching~\cite{deepcache-24, liu2024faster-tgate, encoder-skip-24} methods. In this paper, we mainly focus on training-free acceleration methods that leverages different kinds of feature redundancy throughout the diffusion process. As observed by prior works, adjacent steps exhibits temporal redundancy, happening in layer outputs or even block outputs.
In U-net, the attention module is computed as:
\begin{equation}
    M = \textsf{Softmax}(\frac{Q\cdot K^T}{\sqrt{d}}), O = M \cdot V
\end{equation}
where $Q$ refers to projected features from latent, $K, V$ refers to latent (self-attention) or text embedding (cross-attention).
Recently, Faster Diffusion~\cite{encoder-skip-24} explored the feasibility of skip U-net encoder computation with a delicate skip strategy. 
Applying these acceleration methods to the new paradigm of hybrid inference presents unique challenges that necessitate a reexamination of previous strategies.

%% file: tex/4-design.tex
\section{\name: Design}\label{sec:design}
\begin{figure}[h]
    % \vspace{-0.6cm}
    \centering
    \includegraphics[width=0.6\linewidth]{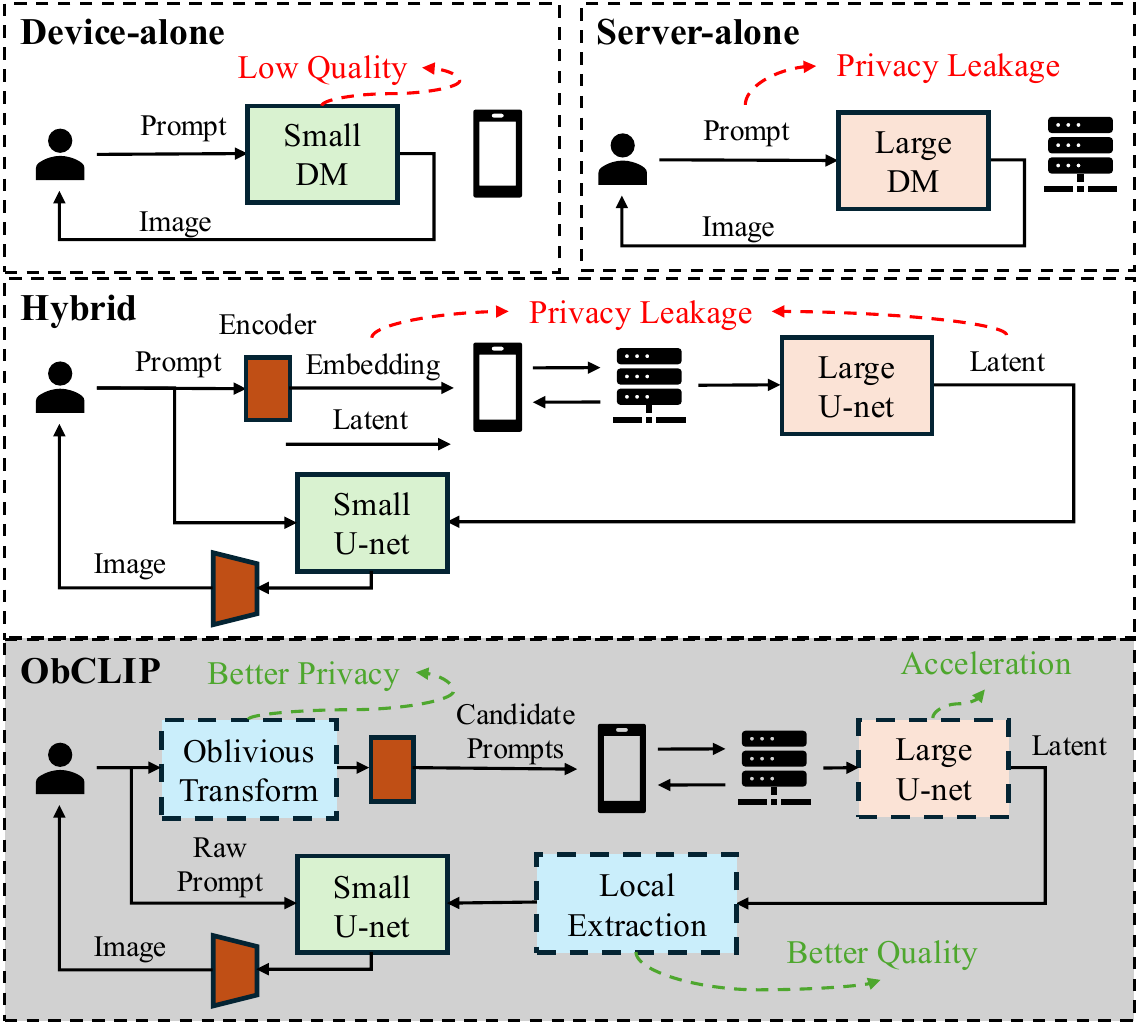}
    \caption{Holistic comparison of existing schemes.}
    \label{fig:overview}
    % \vspace{-0.3cm}
\end{figure}

Inspired by prior works~\cite{ediff-i-22, slim-dpm-23}, which highlight the feasibility of employing a mixture of diffusion models at different stages of the denoising process, we devise \name for an optimal cloud-device hybrid generation scheme with privacy preservation.
As illustrated in Figure~\ref{fig:overview}, the two key components in \name are oblivious transformation and local extraction. By transforming the private prompt into a set of candidate prompts, server only acting as guidance of several initial steps and offloading most of later diffusion steps to device, we achieve the following features: 
1) enhanced image quality by leveraging the on-cloud large-capacity models; 2) strengthened prompt privacy through on‐device computation.
To further lower the server computation cost, we devise several server-side acceleration methods tailored to our scenario.

% \begin{wrapfigure}{r}{0.5\linewidth}
% \end{wrapfigure}

\subsection{Oblivious Cloud-Device Hybrid Image Generation}\label{sec:hybrid}

Before delving into the detailed design of \name, we first answer the two essential research questions.
\begin{itemize}[after=\vspace{-0.5em}]
    \item \textbf{RQ 1}: How to decide the \textit{diffusion allocation strategy} to achieve better utility?
    \item \textbf{RQ 2}: How to \textit{hide the sensitive attributes in a prompt} from the server?
\end{itemize}

\begin{figure*}[ht]
    \centering
    \setlength{\abovecaptionskip}{0pt}
    \setlength{\belowcaptionskip}{0pt}
    \begin{subfigure}[b]{0.42\linewidth}
        \includegraphics[width=0.95\linewidth]{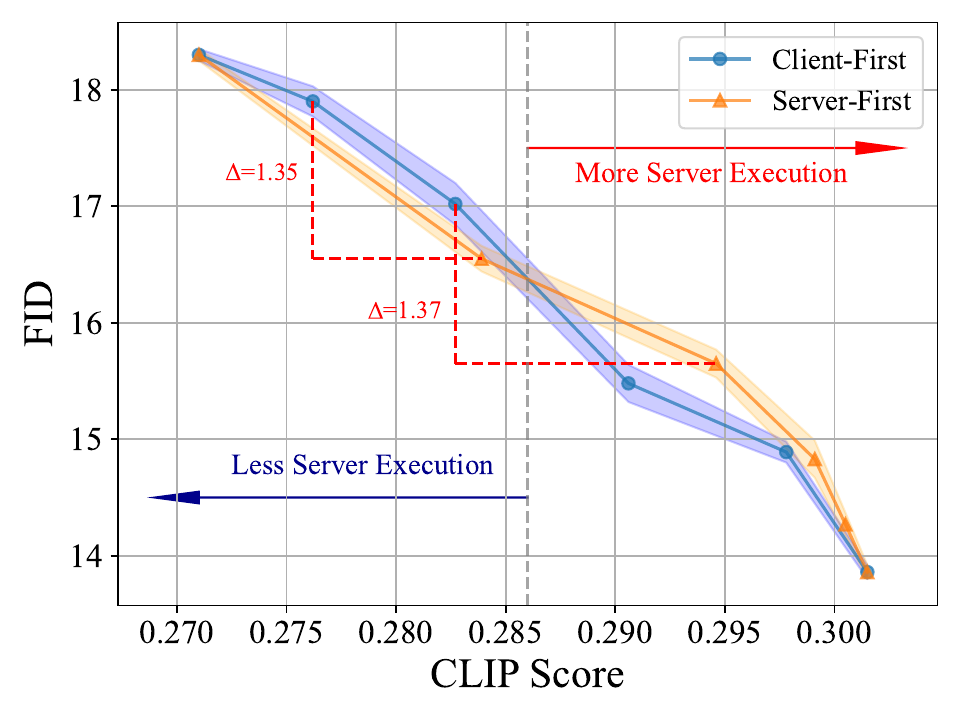}
        \caption{Quantitative quality score. For each line, from left to right, the server execution proportion starts from 0\% to 100\% with interval 20\%.}
        \label{fig:cs-strategy-quantitative}
    \end{subfigure}
    \hfill
    \begin{subfigure}[b]{0.55\linewidth}
        \includegraphics[width=\linewidth]{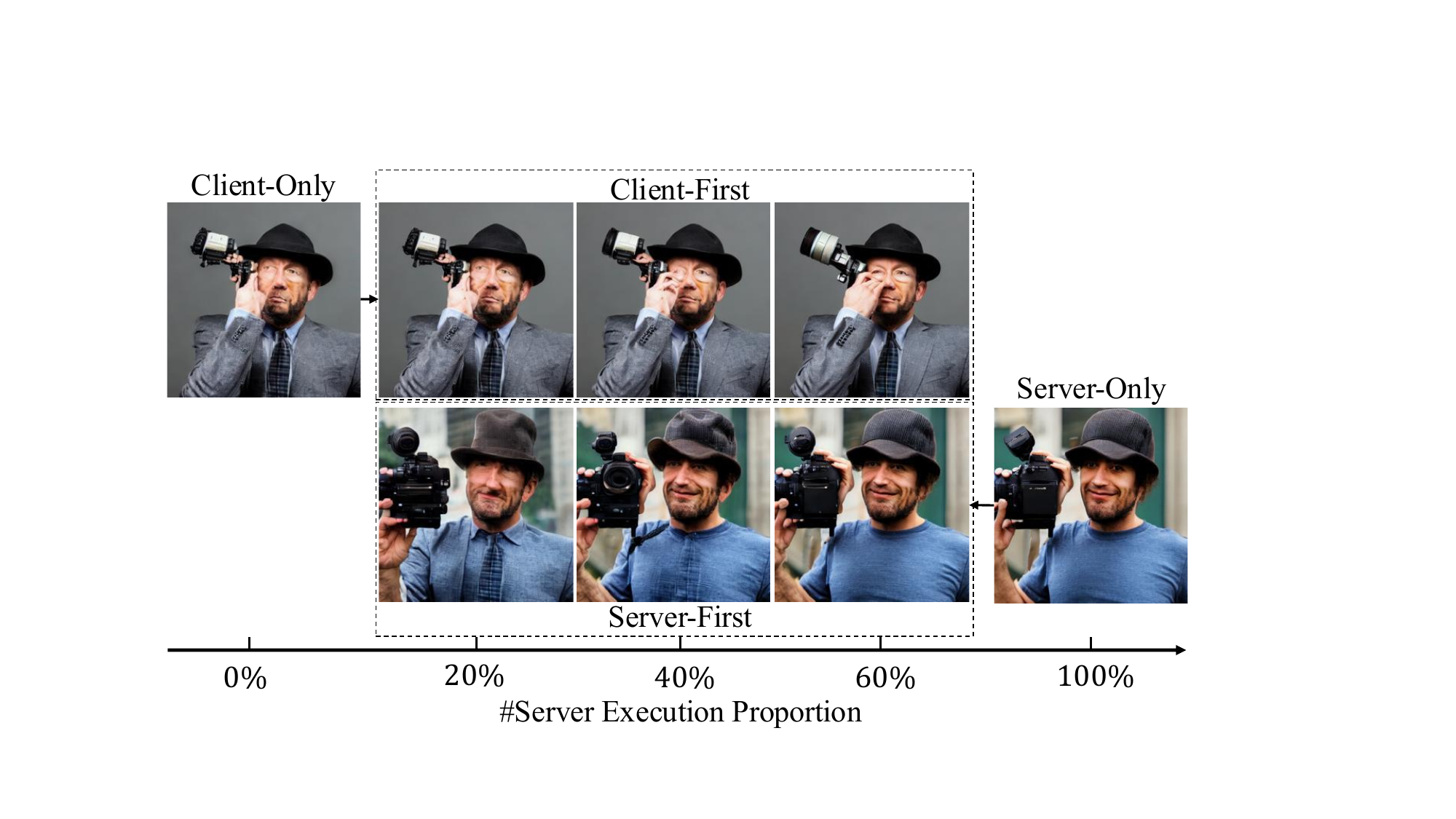}
        \caption{Qualitative quality visualization. Prompt = `A man holding a camera up over his left shoulder.' }
        \label{fig:cs-strategy-qualitative}
    \end{subfigure}
    \caption{Quantitative and qualitative comparison result of using different hybrid execution strategies.}
    \label{fig:cs-strategy}
\end{figure*}

\noindent\textbf{Diffusion Allocation Strategy.}
As an answer to \textbf{RQ 1}, we empirically investigate the impact of different allocation strategies, focusing on two key factors: 1) whether the initial denoising steps are performed by a large server-side model or a small client-side model; and 2) the proportion of diffusion steps distributed between the server and the client.
We employ a 25-step DPM-Solver~\cite{DPM-solver-23} on the SD-v1.4 model and its compressed variant BK-SDM-Small~\cite{bk-sdm-24}, evaluated on the MS-COCO dataset~\cite{mscoco-14}.
We run both quantitative and qualitative analysis by varying server execution proportion.
As shown in Figure~\ref{fig:cs-strategy-quantitative}, when server involvement is limited (0\%$\sim$40\%), allowing the server-side model to handle the initial diffusion steps yields better quantitative performance (i.e., lower FID) compared to the client-first strategy (the performance gap is highlighted with a red dashed line). This observation is consistent with the qualitative results in Figure~\ref{fig:cs-strategy-qualitative}.
As revealed in \cite{liu2024faster-tgate}, the initial steps—referred to as the semantics-planning stage—are crucial for determining global semantic information. Consequently, server-guided semantics planning leads to better image quality.
Based on these findings, we adopt the server-first strategy in our approach. 
Besides, we also control the proportion of server-side execution using the hyper-parameter \textit{switch point} $k$, which serves as a trade-off between utility and efficiency. As $k$ increases, the FID consistently decreases, albeit at the cost of higher server cost.
More comprehensive evaluation are provided in Section~\ref{sec:quantitative}.

\begin{figure}[h] 
  \centering
  \setlength{\abovecaptionskip}{0pt}
   \setlength{\belowcaptionskip}{2pt}
  % \vspace{-0.5cm}
    \includegraphics[width=0.5\linewidth]{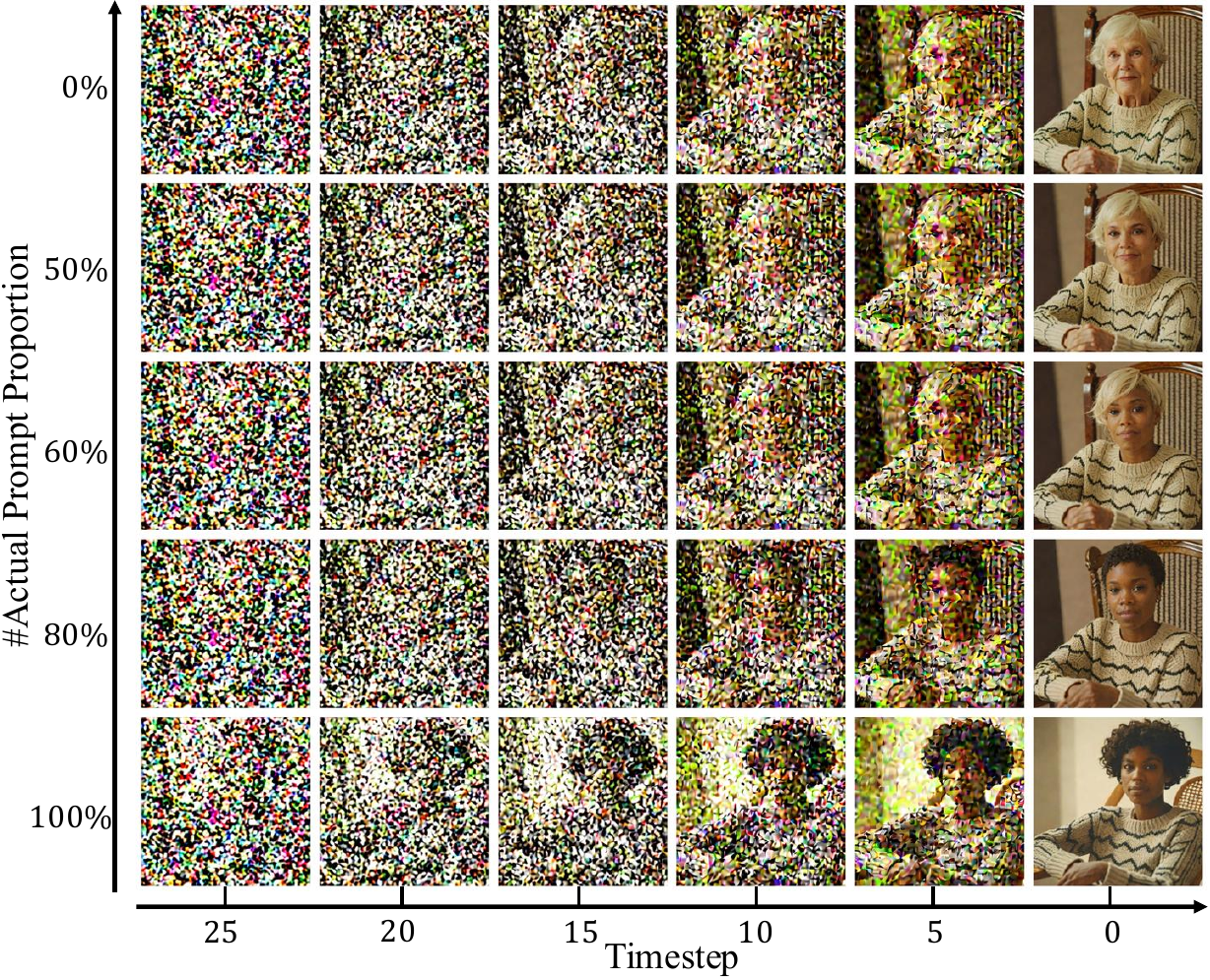}
    \caption{Generated images from different prompt replacement configurations.}
    \label{fig:cs-strategy-rep-vis}
  % \vspace{-0.45cm}
\end{figure}

\noindent\textbf{Oblivious Generation Scheme.}
Regarding \textbf{RQ2}, one straightforward solution is to replace the actual prompt (e.g., ``\textit{portrait of young African woman}'') to a random candidate prompt (e.g., ``\textit{portrait of elderly Caucasian woman}'', semantically-close yet differ in sensitive attributes) to the server, who sends back intermediate denoised latent. The client then use the actual prompt for remaining denoising steps, aiming to \textit{rectify the semantics deviation}. 
We examine the impact of different denoising proportions where the initial steps are conditioned on a candidate prompt, and the remaining steps use the actual prompt as the text condition.
We conduct the analysis using SDXL~\cite{sdxl-24} model with a 25-step DPM-Solver. 
As illustrated in Figure~\ref{fig:cs-strategy-rep-vis}, the intended semantics are accurately captured only when the actual prompt governs more than 80\% of the entire diffusion process. This observation suggests that initial steps are critical for establishing semantic information, making it difficult to correct semantic deviations introduced early on.
Hence, to enhance the text-image semantic alignment, we propose the oblivious hybrid generation scheme, where the client transforms the actual prompt into a set of candidate prompts (including the actual prompt, security analysis in Section~\ref{sec:security}), serving as text conditions during the server-side guidance of intermediate latents. The intended intermediate latents are then retrieved by client for subsequent denoising. To construct the candidate prompt set $\mathcal{P}$, we identify sensitive attributes and traverses their value space as the algorithm in Appendix~\ref{append:hybrid-algo}.

% \begin{wrapfigure}{r}{0.45\textwidth}
% \end{wrapfigure}

Notably, vanilla oblivious generation—where the server performs all the denoising and sends generation images to client—incurs an $N\times$ increase in total computation cost ($N = |\mathcal{P}|$).
While partial denoising in \name effectively reduces server-side computation, the additional overhead introduced by redundant denoising of multiple candidate prompts remains non-negligible. To mitigate this, we introduce server-side acceleration techniques aimed at minimizing such overhead.

\subsection{Server-side Acceleration}\label{sec:server-opt}

\begin{figure}[h]
    \centering
    \setlength{\abovecaptionskip}{5pt}
    \setlength{\belowcaptionskip}{0pt}
    % \vspace{-1.5cm}
    \includegraphics[width=0.6\linewidth]{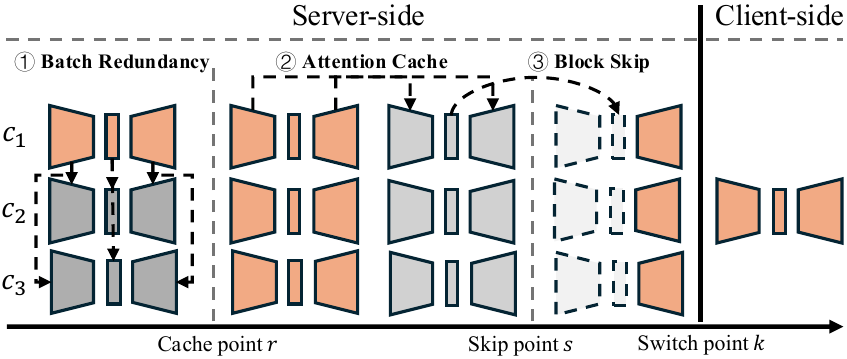}
    \caption{Server-side acceleration. The temporal- and batch- redundancy based caching are controlled by three hyper-parameters: cache point $r$, skip point $s$ and switch point $k$.}
    \label{fig:cache-strategy-vis}
    % \vspace{-0.5cm}
\end{figure}

Conceptually, in \name, we optimize server-side generation efficiency from two perspectives:
1) \textbf{Batch Redundancy}: In oblivious generation, a set of candidate prompts is processed, which differ only in the values of sensitive attributes while sharing most tokens, theoretically leading to similar global semantics, with minor changes on local details.
2) \textbf{Temporal Redundancy}: Due to the inherently sequential nature of the denoising process, intermediate features—such as outputs from attention modules and down/mid blocks—across adjacent timesteps can be cached and reused.
High-level overview of the acceleration scheme is illustrated in Figure~\ref{fig:cache-strategy-vis}.

\subsubsection{Batch Redundancy}
As empirically validated by the visualization of both cross-attention maps (Figure~\ref{fig:cross-attn-vis}) and self-attention maps (deferred to Figure~\ref{fig:self-attn-vis} in Appendix~\ref{append:self-attn-visualization}) that implicitly reflect semantic information for two candidate prompts by varying \textit{gender} and \textit{age} attributes, the global features like background, gesture, etc. are similar, while those sensitive attributes share similar focus areas.
In this regard, we propose to reuse these attention maps across these candidate prompts. Specifically, we only compute the attention map for pivot prompt (with index $i^*$, e.g., the first prompt with $i^*=0$), and reuse these attention maps before \textit{cache point} to lower server-side attention computation as follows:
\begin{align}
    q^*, k^*, V &= \texttt{to\_q}(\mathcal{Q}[i^*]), \texttt{to\_k}(\mathcal{K}[i^*]), \texttt{to\_v}(\mathcal{V})\\
    m^* &= \texttt{get\_attention\_map}(q^*, k^*) \\
    O &= M \cdot V \{M \leftarrow \texttt{broadcast}(m^*)\}
\end{align}
In this case, the computation bottleneck of attention module (i.e., \texttt{to\_q}, \texttt{to\_k}, and Softmax in \texttt{get\_attention\_map}) can be greatly reduced. The detailed algorithm is deferred to Appendix~\ref{append:batch-reuse-algo}. 

\subsubsection{Temporal Redundancy}

\begin{figure*}[ht]
    \centering
    \setlength{\abovecaptionskip}{0pt}
    \setlength{\belowcaptionskip}{0pt}
    \begin{subfigure}[b]{0.22\linewidth}
        \centering
        \includegraphics[width=\linewidth]{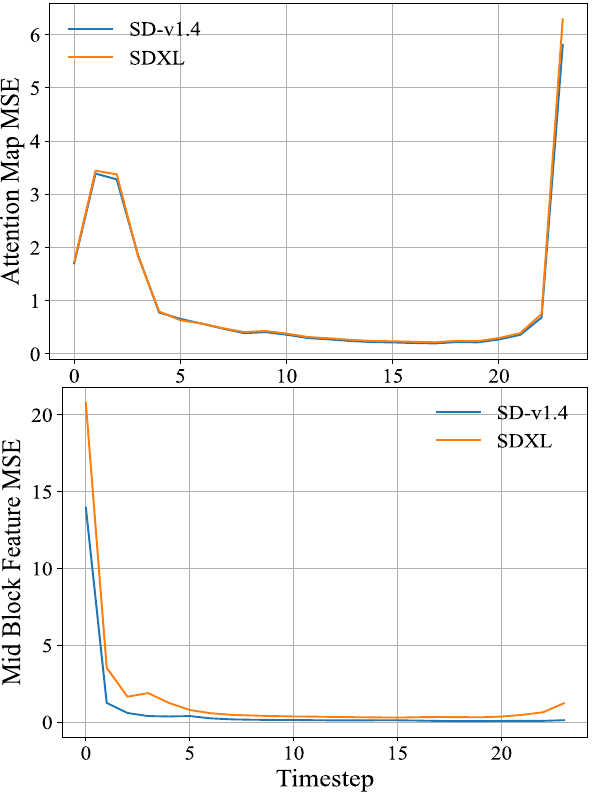}
        \caption{Temporal difference.}
        \label{fig:cache-analysys}
    \end{subfigure}
    \hfill
    \begin{subfigure}[b]{0.77\linewidth}
        \centering
        \includegraphics[width=\linewidth]{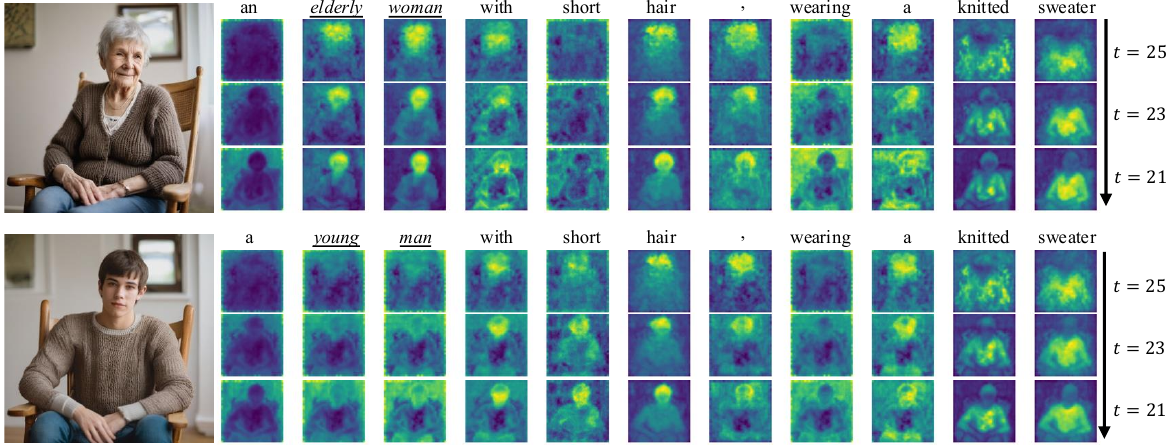}
        \caption{Heatmap of cross-attention maps in two candidate prompts varying \{\textit{gender}, \textit{age}\}.}
        \label{fig:cross-attn-vis}
    \end{subfigure}
    \caption{Temporal and batch redundancy analysis on SD-v1.4 and SDXL models.}
    \label{fig:feature-cache}
\end{figure*}

\noindent\textbf{Attention Cache.}
We motivate this optimization by T-Gate~\cite{liu2024faster-tgate}, which observes that in the early phase (i.e., semantic-planning) conducted on the server side, self-attention makes limited contributions. We thus bypass the self-attention computations in subsequent diffusion steps after certain initial steps (denoted as \textit{cache point} $r$), making use of such temporal redundancy~\cite{liu2024faster-tgate}.
Furthermore, we investigate the evolution of cross-attention map differences across adjacent timesteps. As shown in Figure~\ref{fig:cache-analysys} (top), these differences are substantial during the first 2$\sim$3 steps, but drop significantly and stabilize thereafter. In addition, the cross-attention heatmap visualization in Figure~\ref{fig:cross-attn-vis} shows that the distribution at the \textit{3rd} step is already similar to that at the \textit{5th} step. We thus propose to cache the cross-attention maps after \textit{cache point} as well. Notably, we follow T-Gate to refresh the cache every 5 steps to prevent significant deviations. The attention module is computed as follows:
\begin{equation}
O_{t} = \begin{cases}
 \texttt{Attn}(\mathcal{Q}, \mathcal{K}, \mathcal{V}) &\text{if }  t \leq r \text{ or } (t\mod 5) =0 \\
 O^* &\text{if } t > r \text{ and } (t \mod 5) !=0
\end{cases}
\end{equation}

\noindent\textbf{Block Skip.}
Previous observations~\cite{deepcache-24, encoder-skip-24} have revealed that features from the down-block and mid-block exhibit relatively subtle variations across adjacent timesteps, especially when compared to those from the subsequent up-blocks. 
As illustrated in Figure~\ref{fig:cache-analysys} (bottom), the mid-block outputs show minimal changes during the denoising process—dropping significantly within the first 2$\sim$3 steps and stabilizing after approximately 20\% of total denoising steps.
Motivated by this temporal feature similarity, we propose to skip server-side computation of the down-block and mid-block after a certain timestep (denoted as \textit{skip point} $s$). We follow an intuitive way to select the skip point and defer the parameter configurations to Section~\ref{sec:experiments}. The computation is formulated as follows, where $f_{mid}$ denotes the cached output of mid-block and $\mathcal{P}$ refers to the set of candidate prompts.
\begin{equation}
    z_{t} = \begin{cases}
    (\texttt{Downblock} \circ \texttt{MidBlock} \circ \texttt{Upblock})(z_{t-1}, \mathcal{P}, t)&\text{if } t < s \\
    \texttt{Upblock}(z_{t-1}, f_{mid}, \mathcal{P}, t) &\text{if } t \ge s
\end{cases}
\end{equation}

\subsection{Security Analysis}\label{sec:security}
In \name, the client $\mathcal{C}$ transforms the actual prompt $p^*$ into a set of candidate prompts $\mathcal{P}$, with $|\mathcal{P}| = N$.
We consider the LLM inference service provider as the potential adversary $\mathcal{A}$. We do not consider man-in-the-middle attacks and assume the communication channel is secure.
$\mathcal{A}$ is assumed to be \textit{semi-honest}, meaning the adversary follows the hybrid generation scheme, which is publicly known, but may attempt to extract sensitive information by collecting and analyzing the messages (i.e., $\mathcal{P}$) from the client.
Our scheme shows that $\mathcal{A}$ cannot distinguish the real prompt $p^*$ from $N$ candidates with probability better than $1/N + \lambda$, where $\lambda$ is negligible given no information other than $\mathcal{P}$.
We defer the detailed proof of Theorem~\ref{theorem:oblivious} to Appendix~\ref{append:security}.
\begin{theorem}[Prompt In-distinguishability]\label{theorem:oblivious}
    The oblivious generation scheme is $\lambda$-oblivious if for any probabilistic polynomial-time (PPT) adversary $\mathcal{A}$:
    \begin{equation*}
        |\mathrm{Pr}[A(\mathcal{P} = p^*)] - \frac{1}{N}| \leq \lambda
    \end{equation*}
\end{theorem}

%% file: tex/5-experiments.tex
\section{Experiments}\label{sec:experiments}
\subsection{Experiment Settings}\label{sec:exp-setting}

\noindent\textbf{Models.}
We consider several combinations for hybrid generation. We consider SD-v1.4~\cite{ldm-22}, and its compressed versions BK-SDM-small and BK-SDM-tiny~\cite{bk-sdm-24}. 
We also test finetuned Realistic Vision v4.0~\cite{realisticvision4} and compressed small-sd model~\cite{segmind2023smallsd}.
For high-resolution setting, we consider SDXL~\cite{sdxl-24} and Koala-700m~\cite{koala-nips24}, along with the scalability to step-distilled server model LCM-SDXL~\cite{lcm-23}.

\noindent\textbf{Datasets \& Metrics.}
To evaluate the performance of \name, we adopt two commonly-used datasets: 1) MS-COCO 2014 dataset~\cite{mscoco-14} with a resolution of $512\times512$. We use 30k prompts from its validation split. 2) MJHQ~\cite{mjhq-30k} with a resolution of $1024\times1024$.
For more comprehensive evaluation on oblivious generation, we construct a candidate prompt dataset using 10 templates, like ``\textit{High-quality, face portrait photo of a <age> <ethnicity> <gender>}'' with random fill on these sensitive attributes. 
The detailed construction is provided in Appendix~\ref{append:templates}. 
Regarding image quality, we follow prior works to evaluate the visual quality using Frechet Inception Distance (FID)~\cite{fid-17} and Inception Score (IS)~\cite{is-16}. We assess text-image alignment using CLIP score~\cite{hessel2021clipscore} with CLIP-ViT-g/14 model.
Regarding efficiency, we use Floating-point Operations (FLOPs) and average running time.

\noindent\textbf{Baselines.} 
We compare \name with three lines of works:
1) \textbf{\textit{Standalone Generation}}: The commonly used paradigm where cloud-only or device-only generation is applied. This exhibits either high privacy risk and server cost or low utility;
2) \textbf{\textit{Hybrid SD}}~\cite{hybridsd-24}: This is the first work that proposed cloud-device collaborative generation paradigm. 
3) \textbf{\textit{HE-Diffusion}}~\cite{he-diffusion-24}: 
Last, we compare with cryptographic approach HE-Diffusion in terms of efficiency.
We take the runtime of the SD-v1.4 model as reported in their paper.
We opt for 25-step DPM scheduler (8-step for LCM-SDXL) for all evaluated works. For \name, we mainly adopt two acceleration configurations: 1) switch point $k =5$, cache point $r=3$ and skip point $s=3$; 2) $k=10$, $r=4$ and $s=6$. 
We use $N$ to denote the cardinality of candidate prompt set, i.e., those edited prompts after oblivious transformation. To do so, we use rule-based method and a finetuned distilbert model~\cite{isotonic_distilbert_finetuned_ai4privacy_v2} to detect sensitive attributes. 
We additionally evaluate vanilla oblivious generation (OG), where the cloud alone generates $N$ images without any accelerations. 
All the experiments are conducted on one Ubuntu machine equipped with one Intel Xeon Platinum 8260 CPU, 16GB of RAM and 1 NVIDIA Tesla-V100-SXM2-32GB GPU.

\subsection{Quantitative Results}\label{sec:quantitative}

\noindent\textbf{Results on Candidate Prompt Dataset.}
To begin, we run experiments on Realistic Vision v4.0 and small-sd models on the candidate prompt dataset to evaluate oblivious generation. 
For gender alone, $N=2$. While for multi-attribute combinations, i.e., gender and age, we have $N=2\times3=6$.
We present the results for 1-attribute and 2-attribute oblivious generation in Table~\ref{tab:candidate-prompts}. The results for 3-attribute are deferred to Appendix~\ref{append:3-attribute}.
We here mainly focus on cloud-side latency.
In general, \name shows better performance in the trade-off between generation performance and computation cost, offering flexibility through the acceleration strategies, e.g., $k$.
In terms of generation performance, \name achieves comparable and even better FID and IS compared to Realistic Vision v4.0 when $k=10$ (e.g., FID drops from 113.39 to 109.76 when $N=6$). Even when we use a more aggressive $k=5$, we still achieve comparable FID and IS while much lower latency (e.g., FID increases from 113.39 to 113.92 with latency decreases from 1.12s to 0.98s).
In terms of latency, compared to Hybrid SD, \name incurs about $N\times$ FLOPs due to the oblivious generation. However, with the cache and reuse accelerations enabled, the latency of \name is nearly reduced by 50\%, which is comparable to Hybrid SD when $N=2$ (e.g., latency increases from 0.55s to 0.57s when $k=10$) and about $3\times$ slower when $N=6$. Given the strong privacy protection brought by \name, such computation cost is significantly reduced. Especially,\name is orders of magnitude faster than HE-Diffusion and about $4.4\sim7.6\times$ faster than vanilla OG.
The results validate the effectiveness of \name, which offers rigorous privacy and better generation performance (measured by FID and CLIP scores), with only a marginal increase in latency.

\begin{table}[ht]
    \centering
    \caption{Multi-Attribute \name on candidate prompt dataset. For FLOPs, we use $a (+b)$, where $a$ and $b$ refer to cloud/device computations. For latency, we only measure the cloud-side runtime.}
    \scalebox{0.7}{
    \begin{tabular}{@{}l|ccccc|ccccc@{}}
\toprule
\multicolumn{1}{c|}{}                                             & \multicolumn{5}{c|}{\textbf{1-Attribute (gender, $N=2$)}}                                                                                                                                                                                         & \multicolumn{5}{c}{\textbf{2-Attribute (gender + age, $N=6$)}}                                                                                                                                                                                    \\ \cmidrule(l){2-11} 
\multicolumn{1}{c|}{\multirow{-2}{*}{\textbf{Generation Method}}} & \multicolumn{1}{c|}{\textbf{FID $\downarrow$}}      & \multicolumn{1}{c|}{\textbf{IS $\uparrow$}}       & \multicolumn{1}{c|}{\textbf{CLIP $\uparrow$}}       & \multicolumn{1}{c|}{\textbf{FLOPs (T)}}                    & \textbf{Latency (s)} & \multicolumn{1}{c|}{\textbf{FID $\downarrow$}}      & \multicolumn{1}{c|}{\textbf{IS $\uparrow$}}       & \multicolumn{1}{c|}{\textbf{CLIP $\uparrow$}}       & \multicolumn{1}{c|}{\textbf{FLOPs (T)}}                    & \textbf{Latency (s)} \\ \midrule
Realistic Vision v4.0                                             & \multicolumn{1}{c|}{113.45}                         & \multicolumn{1}{c|}{4.69}                         & \multicolumn{1}{c|}{0.3322}                         & \multicolumn{1}{c|}{18.53 (+0)}                            & 1.12                 & \multicolumn{1}{c|}{113.39}                         & \multicolumn{1}{c|}{5.32}                         & \multicolumn{1}{c|}{0.3215}                         & \multicolumn{1}{c|}{18.53 (+0)}                            & 1.12                 \\
small-sd                                                          & \multicolumn{1}{c|}{128.87}                         & \multicolumn{1}{c|}{5.04}                         & \multicolumn{1}{c|}{0.3051}                         & \multicolumn{1}{c|}{0 (+11.20)}                            & 0.78                 & \multicolumn{1}{c|}{118.19}                         & \multicolumn{1}{c|}{5.11}                         & \multicolumn{1}{c|}{0.2980}                         & \multicolumn{1}{c|}{0 (+11.20)}                            & 0.78                 \\ \midrule
Vanilla OG                                                        & \multicolumn{1}{c|}{113.45}                         & \multicolumn{1}{c|}{4.69}                         & \multicolumn{1}{c|}{0.3322}                         & \multicolumn{1}{c|}{37.06 (+0)}                            & 2.51                 & \multicolumn{1}{c|}{113.39}                         & \multicolumn{1}{c|}{5.32}                         & \multicolumn{1}{c|}{0.3215}                         & \multicolumn{1}{c|}{111.18 (+0)}                           & 7.47                 \\ \cmidrule(l){2-11} 
HE-Diffusion                                                      & \multicolumn{4}{c|}{-}                                                                                                                                                                                                     & \textgreater 106     & \multicolumn{4}{c|}{-}                                                                                                                                                                                                     & \textgreater 106     \\ \midrule
Hybrid SD ($k=10$)                                                & \multicolumn{1}{c|}{117.18}                         & \multicolumn{1}{c|}{4.96}                         & \multicolumn{1}{c|}{0.3215}                         & \multicolumn{1}{c|}{7.41 (+6.54)}                          & 0.55                 & \multicolumn{1}{c|}{114.05}                         & \multicolumn{1}{c|}{5.02}                         & \multicolumn{1}{c|}{0.3226}                         & \multicolumn{1}{c|}{7.41 (+6.54)}                          & 0.55                 \\
$\name (k=10)$                                                    & \multicolumn{1}{c|}{117.18}                         & \multicolumn{1}{c|}{4.96}                         & \multicolumn{1}{c|}{0.3215}                         & \multicolumn{1}{c|}{14.82 (+6.54)}                         & 0.97                 & \multicolumn{1}{c|}{114.05}                         & \multicolumn{1}{c|}{5.02}                         & \multicolumn{1}{c|}{0.3226}                         & \multicolumn{1}{c|}{44.46 (+6.54)}                         & 2.90                 \\
$\enspace\textbf{+}$ cache                                        & \multicolumn{1}{c|}{118.59}                         & \multicolumn{1}{c|}{4.99}                         & \multicolumn{1}{c|}{0.3168}                         & \multicolumn{1}{c|}{12.26 (+6.54)}                         & 0.62                 & \multicolumn{1}{c|}{115.65}                         & \multicolumn{1}{c|}{5.02}                         & \multicolumn{1}{c|}{0.3174}                         & \multicolumn{1}{c|}{36.76 (+6.54)}                         & 1.85                 \\
\rowcolor[HTML]{C0C0C0} 
$\enspace\textbf{+}$ reuse                                        & \multicolumn{1}{c|}{\cellcolor[HTML]{C0C0C0}114.26} & \multicolumn{1}{c|}{\cellcolor[HTML]{C0C0C0}4.82} & \multicolumn{1}{c|}{\cellcolor[HTML]{C0C0C0}0.3167} & \multicolumn{1}{c|}{\cellcolor[HTML]{C0C0C0}11.48 (+6.54)} & 0.57                 & \multicolumn{1}{c|}{\cellcolor[HTML]{C0C0C0}109.76} & \multicolumn{1}{c|}{\cellcolor[HTML]{C0C0C0}4.94} & \multicolumn{1}{c|}{\cellcolor[HTML]{C0C0C0}0.3152} & \multicolumn{1}{c|}{\cellcolor[HTML]{C0C0C0}33.28 (+6.54)} & 1.55                 \\ \midrule
Hybrid SD ($k=5$)                                                 & \multicolumn{1}{c|}{119.31}                         & \multicolumn{1}{c|}{4.99}                         & \multicolumn{1}{c|}{0.3107}                         & \multicolumn{1}{c|}{3.71 (+8.96)}                          & 0.28                 & \multicolumn{1}{c|}{116.15}                         & \multicolumn{1}{c|}{5.05}                         & \multicolumn{1}{c|}{0.3117}                         & \multicolumn{1}{c|}{3.71 (+8.96)}                          & 0.28                 \\
$\name (k=5)$                                                     & \multicolumn{1}{c|}{119.31}                         & \multicolumn{1}{c|}{4.99}                         & \multicolumn{1}{c|}{0.3107}                         & \multicolumn{1}{c|}{7.41 (+8.96)}                          & 0.49                 & \multicolumn{1}{c|}{116.15}                         & \multicolumn{1}{c|}{5.05}                         & \multicolumn{1}{c|}{0.3117}                         & \multicolumn{1}{c|}{22.23 (+8.96)}                         & 1.48                 \\
$\enspace\textbf{+}$ cache                                        & \multicolumn{1}{c|}{120.44}                         & \multicolumn{1}{c|}{4.88}                         & \multicolumn{1}{c|}{0.3079}                         & \multicolumn{1}{c|}{6.13 (+8.96)}                          & 0.38                 & \multicolumn{1}{c|}{117.29}                         & \multicolumn{1}{c|}{5.00}                         & \multicolumn{1}{c|}{0.3091}                         & \multicolumn{1}{c|}{18.38 (+8.96)}                         & 1.12                 \\
\rowcolor[HTML]{9B9B9B} 
$\enspace\textbf{+}$ reuse                                        & \multicolumn{1}{c|}{\cellcolor[HTML]{9B9B9B}118.36} & \multicolumn{1}{c|}{\cellcolor[HTML]{9B9B9B}4.98} & \multicolumn{1}{c|}{\cellcolor[HTML]{9B9B9B}0.3077} & \multicolumn{1}{c|}{\cellcolor[HTML]{9B9B9B}5.74 (+8.96)}  & 0.33                 & \multicolumn{1}{c|}{\cellcolor[HTML]{9B9B9B}113.92} & \multicolumn{1}{c|}{\cellcolor[HTML]{9B9B9B}4.87} & \multicolumn{1}{c|}{\cellcolor[HTML]{9B9B9B}0.3076} & \multicolumn{1}{c|}{\cellcolor[HTML]{9B9B9B}16.64 (+8.96)} & 0.98                 \\ \bottomrule
\end{tabular}
}
    \label{tab:candidate-prompts}
\end{table}

\noindent\textbf{Results on Real-world Datasets.}
Table~\ref{tab:sd-mscoco} and Table~\ref{tab:mjhq} (upper part) present the results for MS-COCO and MJHQ datasets. For numbers marked with $^*$, the total FLOPs should be multiplied by $N$.
Compared to distilled models, \name considerably reduces FLOPs while achieving better image fidelity and stronger semantic alignment between image and prompt across various configurations. 
When compared to base models, \name offers substantial FLOPs reduction at the cost of a slightly higher FID.
For instance, on SD and BK-SDM-small models, the FID of \name ($k=10$, with cache) increases from 13.86 to 15.73 with a reduction of FLOPs from 18.53T to $5.84^*$T.
Similarly, on SDXL and Koala-700m models, \name ($k=10$, with cache) achieves a FID of 30.79, which is comparable to SDXL’s 30.67, while reducing the FLOPs from 159.35T to $45.11^*$T — even lower when $N < 4$.
The trade-off between performance and latency can also be controlled via $k$: a larger $k$ yields better image fidelity at the expense of higher FLOPs. 

\begin{table}[htbp]
  \noindent
  \begin{minipage}[t]{0.49\textwidth}
    \centering
    \caption{SD-v1.4 and BK-SDM-\{small, tiny\} on 30k MS-COCO dataset.}
    \scalebox{0.72}{
    \begin{tabular}{l|c|c|c|c}
\toprule
\textbf{Generation Method} & \textbf{FID $\downarrow$} & \textbf{IS $\uparrow$} & \textbf{CLIP $\uparrow$} & \textbf{FLOPs (T)} \\ \midrule
SD-v1.4                    & 13.86                     & 37.75                  & 0.3015                   & 18.53              \\ \midrule
BK-SDM-small               & 18.30                     & 31.73                  & 0.2710                   & 10.90              \\
$\name (k=10)$             & 15.65                     & 36.72                  & 0.2946                   & $7.41^*$           \\
\rowcolor[HTML]{C0C0C0} 
$\enspace\textbf{+}$ cache & 15.73                     & 33.62                  & 0.2865                   & $5.84^*$           \\
$\name (k=5)$              & 16.55                     & 33.95                  & 0.2839                   & $3.71^*$           \\
\rowcolor[HTML]{9B9B9B} 
$\enspace\textbf{+}$ cache & 16.45                     & 33.36                  & 0.2833                   & $3.06^*$           \\ \midrule
BK-SDM-tiny                & 18.30                     & 29.94                  & 0.2681                   & 10.25              \\
$\name (k=10)$             & 15.86                     & 35.54                  & 0.2936                   & $7.41^*$           \\
\rowcolor[HTML]{C0C0C0} 
$\enspace\textbf{+}$ cache & 16.44                     & 32.80                  & 0.2887                   & $5.84^*$           \\
$\name (k=5)$              & 16.87                     & 32.73                  & 0.2812                   & $3.71^*$           \\
\rowcolor[HTML]{9B9B9B} 
$\enspace\textbf{+}$ cache & 17.14                     & 31.84                  & 0.2854                   & $3.06^*$           \\ \bottomrule
\end{tabular}
}
    \label{tab:sd-mscoco}
  \end{minipage}\hfill  % 在两列间自动拉开空白
  \begin{minipage}[t]{0.49\textwidth}
    \centering
    \caption{\{SDXL, LCM-SDXL\} and Koala-700m on 5k MJHQ dataset. $\triangle t$ denotes timestep shift.}
    \scalebox{0.72}{
    \begin{tabular}{@{}l|c|c|c|c@{}}
\toprule
\textbf{Generation Method}                                           & \textbf{FID $\downarrow$}     & \textbf{IS $\uparrow$}        & \textbf{CLIP $\uparrow$}       & \textbf{FLOPs (T)}          \\ \midrule
SDXL                                                                 & 30.67                         & 26.31                         & 0.3464                         & 159.35                      \\
koala-700m                                                           & 36.11                         & 22.06                         & 0.3263                         & 58.85                       \\ \midrule
$\name (k=10)$                                                       & 31.92                         & 24.56                         & 0.3389                         & $63.74^*$                   \\
\rowcolor[HTML]{C0C0C0} 
$\enspace\textbf{+}$ cache                                           & 30.79                         & 24.52                         & 0.3320                         & $45.11^*$                   \\
$\name (k=5)$                                                        & 32.42                         & 23.70                         & 0.3337                         & $31.87^*$                   \\
\rowcolor[HTML]{9B9B9B} 
$\enspace\textbf{+}$ cache                                           & 31.97                         & 23.61                         & 0.3317                         & $24.41^*$                   \\ \midrule
LCM-SDXL                                                             & 33.25                         & 28.22                         & 0.3296                         & 50.99                       \\
$\name$ ($k$=4,$\triangle t$=8)                                      & 33.92                         & 27.55                         & 0.3315                         & $25.50^*$                   \\ \midrule
\cellcolor[HTML]{9B9B9B}$\enspace\textbf{+}$ cache $(\triangle t=8)$ & \cellcolor[HTML]{9B9B9B}34.12 & \cellcolor[HTML]{9B9B9B}27.15 & \cellcolor[HTML]{9B9B9B}0.3311 &                             \\
$\enspace\textbf{+}$ cache $(\triangle t=10)$                        & 45.51                         & 22.64                         & 0.3293                         &                             \\
$\enspace\textbf{+}$ cache $(\triangle t=4)$                         & 39.07                         & 22.73                         & 0.3166                         &                             \\
$\enspace\textbf{+}$ cache $(\triangle t=0)$                         & 51.85                         & 16.94                         & 0.2936                         & \multirow{-4}{*}{$21.78^*$} \\ \bottomrule
\end{tabular}
}
    \label{tab:mjhq}
  \end{minipage}
\end{table}

\noindent\textbf{Scalability to Distilled Models.}
We hereby further explore the scalability of \name to step-distilled cloud models.
Note that the LCM-SDXL uses a 8-step scheduler, while Koala-700m uses standard 25-step scheduler. We adopt $k=4, r=2$ for acceleration.
Besides, to align the denoising timesteps between cloud and device, we propose to apply a timestep shift as $t_{device} = t_{cloud} + \triangle t$. 
A smaller or larger $\triangle t$ introduces incompatible denoising scales.
As depicted in the bottom of Table~\ref{tab:mjhq}, $\triangle t = 8$ yields the best performance (marked in gray). We provide visual examples in Appendix~\ref{append:vis-lcm-sdxl}.
Compared to the standalone LCM-SDXL, \name (without cache) achieves a 0.67 drop in FID, a slightly better CLIP score, and reduces FLOPs from 50.99 to $25.50^*$. With cache enabled, the FLOPs are further reduced by approximately 15\%, with only a marginal FID drop of 0.2.

\noindent\textbf{Additional Overhead.}
In \name, we introduce two additional operations: 1) device-side oblivious transformation; 2) cloud-to-device latent transmission. The runtime for detecting sensitive attributes using the fine-tuned DistilBERT model on a single V100 GPU is about 6.37ms. According to the report~\footnote{\url{https://machinelearning.apple.com/research/neural-engine-transformers}}, the latency on iPhone 13 Pro is below 10ms. For cloud-to-device transmission, the noise latent is sized at $4\times \textsf{res} \times \textsf{res} \times N $.
Taking SD-v1.4 as an example, where $\textsf{res}=64$, the total data in FP16 precision is $\sim 32N$ KB. Considering the average WiFi bandwidth of 18.88Mbps~\cite{wifi-bandwidth}, the total transmission time is around $0.013 N$ seconds. Even for $N=30$, the transmission time is approximately 0.39 seconds, demonstrating an acceptable overhead. 
We note that private information retrieval (PIR)~\cite{SealPIR-18, YPIR-24} can also be employed for this transmission, achieving a communication size of $\mathcal{O}(\log^2N)$ or $\mathcal{O}(\sqrt{N})$, albeit at the expense of $\mathcal{O}(N)$ server computation. This approach may be preferable when the bandwidth is limited.
Due to page limitation, we provide a more comprehensive efficiency evaluation and detailed efficiency improvement breakdown in Appendix~\ref{append:efficiency}$\sim$\ref{append:efficiency-breakdown}.

\subsection{Qualitative Results}\label{sec:qualitative}

\begin{figure*}[h]
    \centering
    \setlength{\abovecaptionskip}{0pt}
    \setlength{\belowcaptionskip}{0pt}
    \begin{subfigure}[b]{0.48\linewidth}
        \centering
        \includegraphics[width=0.95\linewidth]{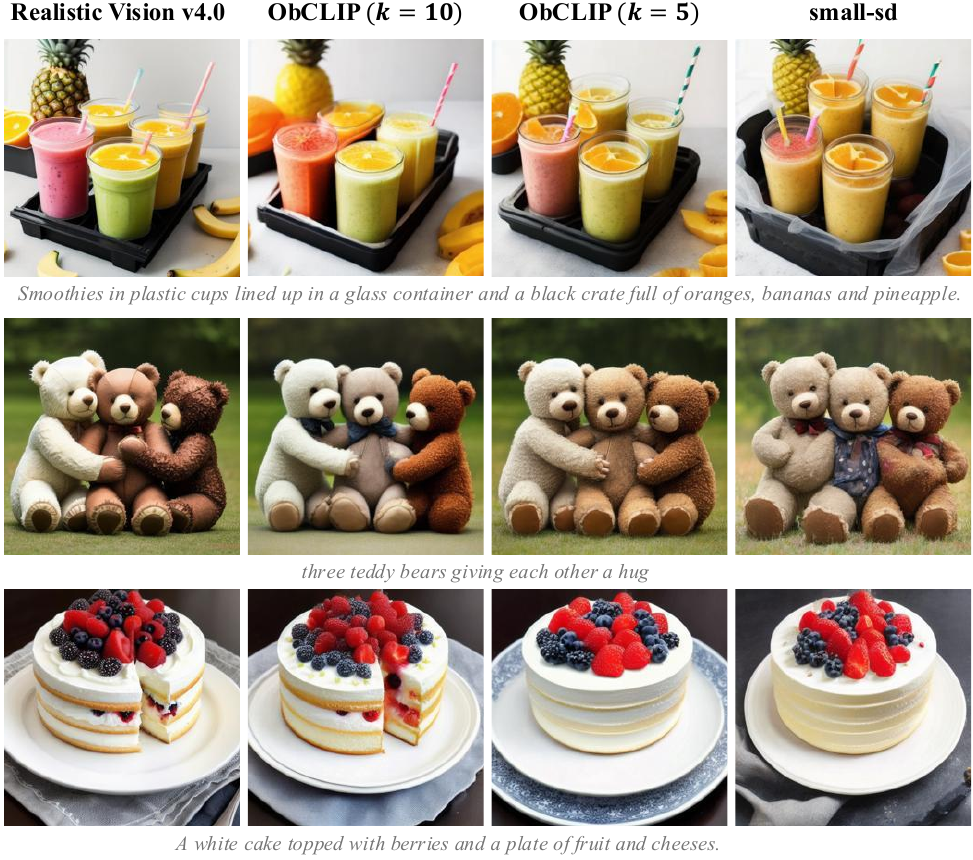}
        \caption{Results on Realistic Vision v4.0 and small-sd.}
        \label{fig:qualitative-realistic}
    \end{subfigure}
    \hfill
    \begin{subfigure}[b]{0.45\linewidth}
        \centering
        \includegraphics[width=0.95\linewidth]{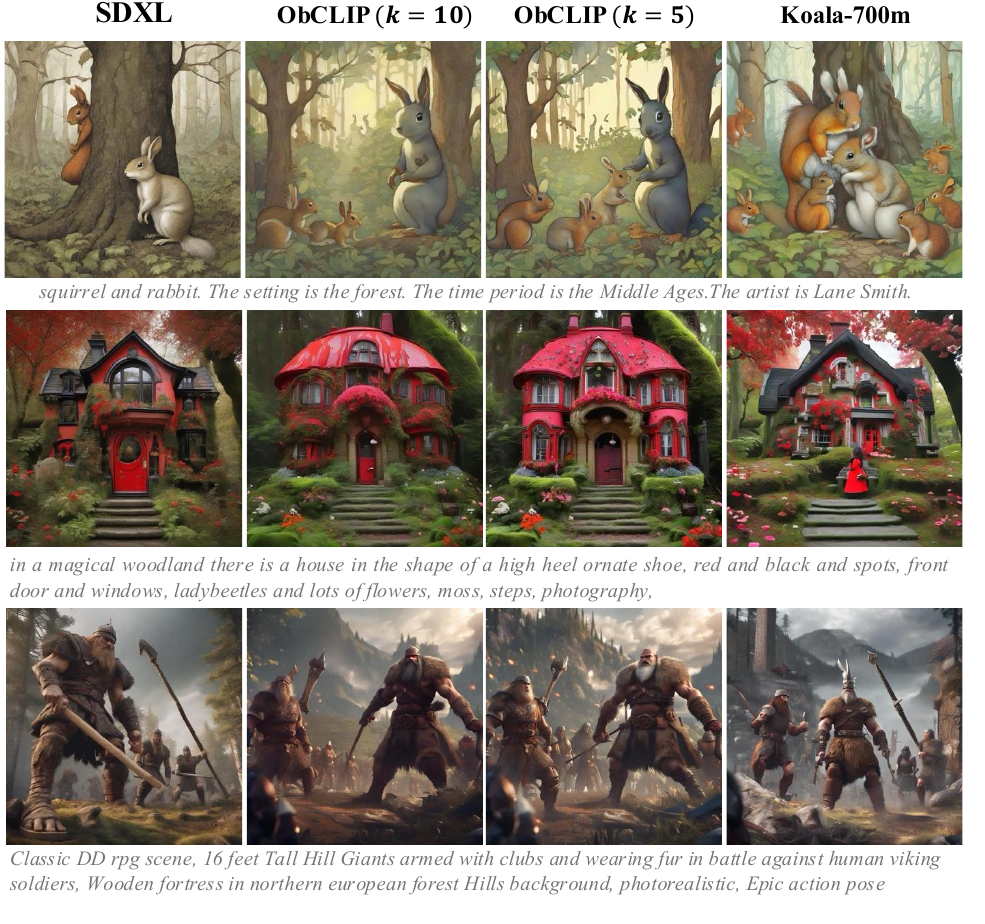}
        \caption{Results on SDXL and Koala-700m.}
        \label{fig:qualitative-sdxl}
    \end{subfigure}
    \caption{Images generated by cloud (left), device (right) and \name (middle) with different $k$.}
    \label{fig:qualitative-vis}
\end{figure*}

\noindent\textbf{Effect of $k$.} Figure~\ref{fig:qualitative-realistic} and Figure~\ref{fig:qualitative-sdxl} show the samples generated using base models and \name (with cache) with different $k$. 
One observation is that with the guidance of large cloud models, \name achieves better semantic alignment and finer local details. 
For example, Koala-700m mistakenly generates a girl for the second prompt, while small-sd generates three overlapping teddy bears.
Besides, a larger $k$ exhibits better visual quality—closer to that of large cloud models—which is consistent with our previous quantitative findings.

\begin{figure*}[ht]
    \centering
    \setlength{\abovecaptionskip}{0pt}
    \setlength{\belowcaptionskip}{0pt}
    \begin{subfigure}[b]{0.44\linewidth}
        \includegraphics[width=1.0\linewidth]{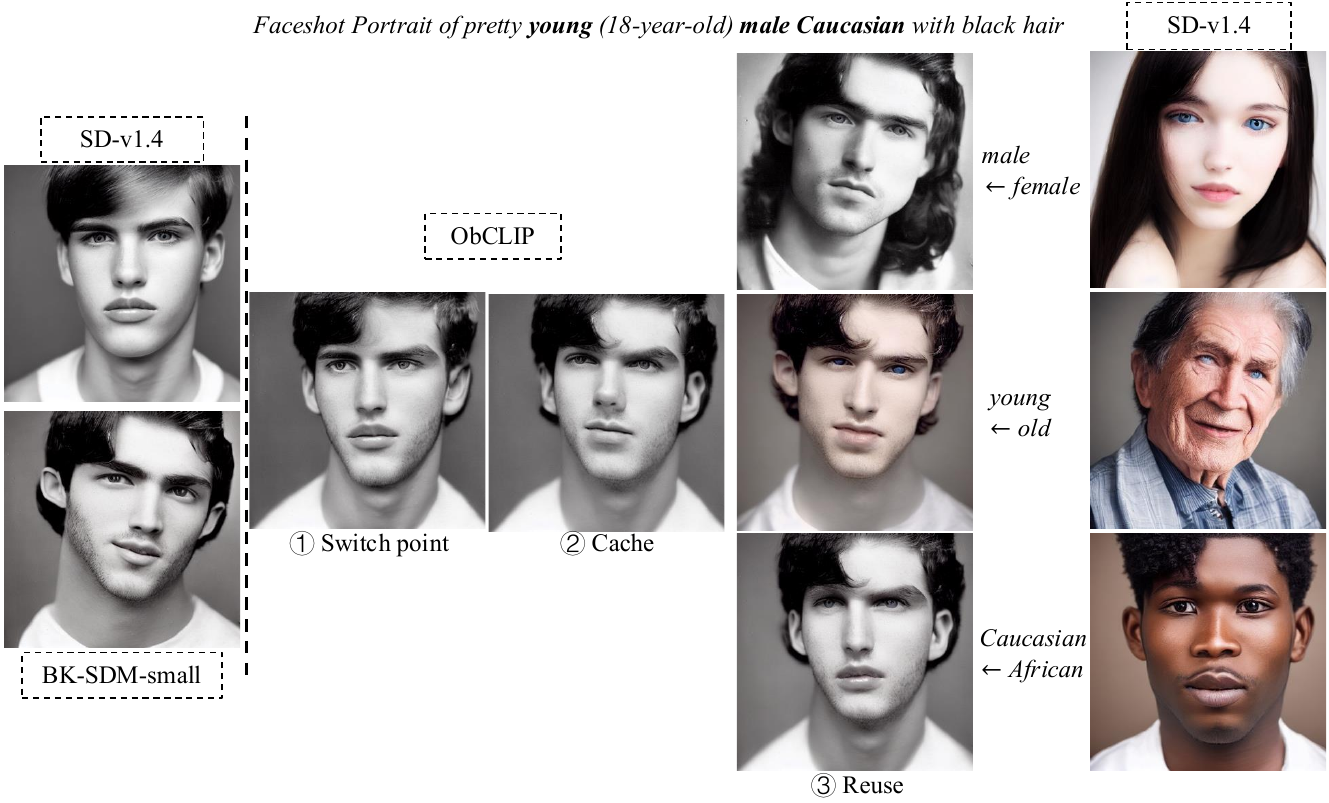}
    \caption{SD-v1.4 and BK-SDM-small}
    \label{fig:vis_sd-batch-reuse}
    \end{subfigure}
    \hfill
    \begin{subfigure}[b]{0.5\linewidth}
        \includegraphics[width=1.\linewidth]{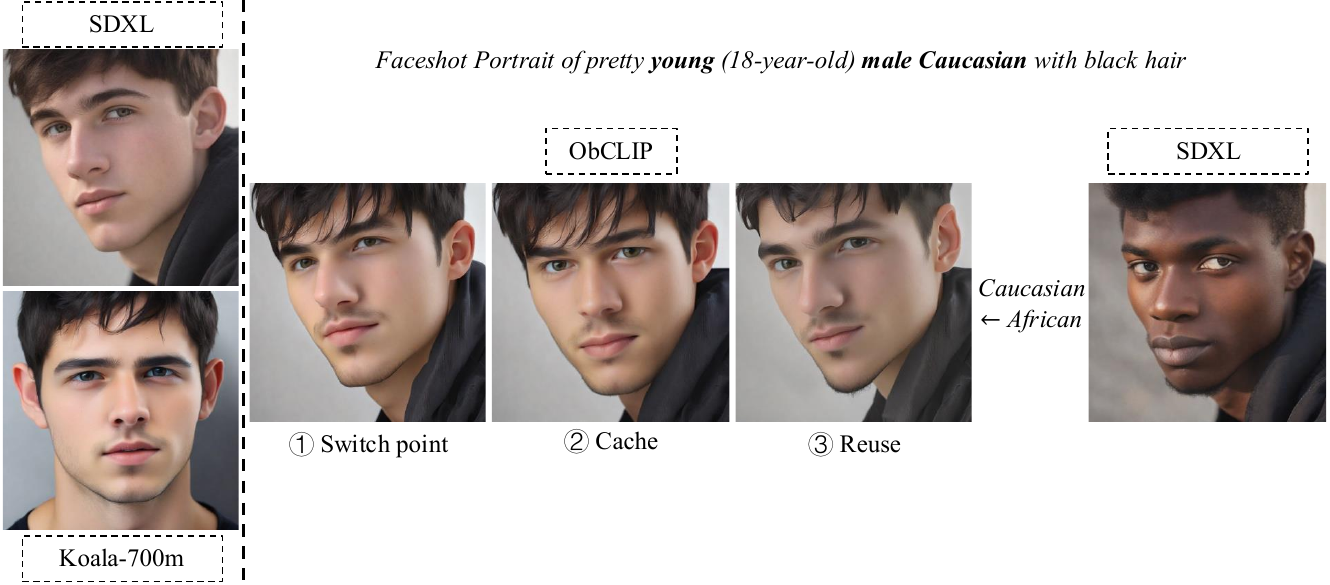}
    \caption{SDXL and Koala-700m}
    \label{fig:vis-sdxl-batch-reuse}
    \end{subfigure}
    \caption{Qualitative visual results when using different server-side acceleration strategies.}
    \label{fig:vis-work}
\end{figure*}

\noindent\textbf{Effect of batch reuse.}
Furthermore, we present the samples generated with different acceleration strategies in Figure~\ref{fig:vis-work}. 
Specifically, the reuse across different gender, age, and ethnicity attributes results in strong semantic alignment, even when the attributes are contradictory (e.g., from female to male).
Global structural information unrelated to the sensitive attributes—such as hairstyle and gesture—are well preserved, while the transformation to the target attribute is effectively accomplished.
For instance, reusing female-associated attention maps in male image generation yields a male with long hair, illustrating both structural consistency and effective attribute modification.

%% file: tex/6-conclusion.tex
\section{Conclusion}\label{sec:conclusion}
In this paper, we propose an oblivious cloud-device hybrid image generation scheme, acting as a plug-and-play safeguard to ML inference services, to provide rigorous prompt privacy, with better utility against on-device generation and only slightly increased server computation cost.
Extensive experiments across multiple datasets demonstrate that \name provides comparable utility to cloud models with slightly increased server cost compared to non-private baselines.

\noindent\textbf{Limitations.}
Despite the significant efficiency improvement over private baselines, a limitation lies in the inherent nature of oblivious generation, which leads to a sub-linear increase in computation relative to the number of candidate prompts. A potential mitigation is to achieve statistical indistinguishability, thereby reducing the effective size of candidate prompts.
This can be achieved by differential privacy-based top-$k$ selection of these candidate prompts. By choosing appropriate privacy budget, we can achieve measurable privacy, while providing a much better efficiency.
In the future, we also plan to extend our approach to the image-to-image generation domain, where inputs include not only text prompts but also real reference images—such as human faces—that carry more sensitive information.

%% file: tex/7-checklist.tex
\newpage
\section*{NeurIPS Paper Checklist}

%%% BEGIN INSTRUCTIONS %%%

%%% END INSTRUCTIONS %%%

\begin{enumerate}

\item {\bf Claims}
    \item[] Question: Do the main claims made in the abstract and introduction accurately reflect the paper's contributions and scope?
    \item[] Answer: \answerYes{}
    \item[] Justification: We provide detailed application scenario and main contributions in the abstract and Section~\ref{sec:intro}. The experiments in Section~\ref{sec:experiments} also exhibit the effectiveness of our work. 
    \item[] Guidelines:
    \begin{itemize}
        \item The answer NA means that the abstract and introduction do not include the claims made in the paper.
        \item The abstract and/or introduction should clearly state the claims made, including the contributions made in the paper and important assumptions and limitations. A No or NA answer to this question will not be perceived well by the reviewers. 
        \item The claims made should match theoretical and experimental results, and reflect how much the results can be expected to generalize to other settings. 
        \item It is fine to include aspirational goals as motivation as long as it is clear that these goals are not attained by the paper. 
    \end{itemize}

\item {\bf Limitations}
    \item[] Question: Does the paper discuss the limitations of the work performed by the authors?
    \item[] Answer: \answerYes{}
    \item[] Justification: We discuss the primary limitations of this work in Section~\ref{sec:conclusion}.
    \item[] Guidelines:
    \begin{itemize}
        \item The answer NA means that the paper has no limitation while the answer No means that the paper has limitations, but those are not discussed in the paper. 
        \item The authors are encouraged to create a separate "Limitations" section in their paper.
        \item The paper should point out any strong assumptions and how robust the results are to violations of these assumptions (e.g., independence assumptions, noiseless settings, model well-specification, asymptotic approximations only holding locally). The authors should reflect on how these assumptions might be violated in practice and what the implications would be.
        \item The authors should reflect on the scope of the claims made, e.g., if the approach was only tested on a few datasets or with a few runs. In general, empirical results often depend on implicit assumptions, which should be articulated.
        \item The authors should reflect on the factors that influence the performance of the approach. For example, a facial recognition algorithm may perform poorly when image resolution is low or images are taken in low lighting. Or a speech-to-text system might not be used reliably to provide closed captions for online lectures because it fails to handle technical jargon.
        \item The authors should discuss the computational efficiency of the proposed algorithms and how they scale with dataset size.
        \item If applicable, the authors should discuss possible limitations of their approach to address problems of privacy and fairness.
        \item While the authors might fear that complete honesty about limitations might be used by reviewers as grounds for rejection, a worse outcome might be that reviewers discover limitations that aren't acknowledged in the paper. The authors should use their best judgment and recognize that individual actions in favor of transparency play an important role in developing norms that preserve the integrity of the community. Reviewers will be specifically instructed to not penalize honesty concerning limitations.
    \end{itemize}

\item {\bf Theory assumptions and proofs}
    \item[] Question: For each theoretical result, does the paper provide the full set of assumptions and a complete (and correct) proof?
    \item[] Answer: \answerYes{}
    \item[] Justification: We provide a detailed security analysis in Section~\ref{sec:security} and proof in the Appendix~\ref{append:security}.
    \item[] Guidelines:
    \begin{itemize}
        \item The answer NA means that the paper does not include theoretical results. 
        \item All the theorems, formulas, and proofs in the paper should be numbered and cross-referenced.
        \item All assumptions should be clearly stated or referenced in the statement of any theorems.
        \item The proofs can either appear in the main paper or the supplemental material, but if they appear in the supplemental material, the authors are encouraged to provide a short proof sketch to provide intuition. 
        \item Inversely, any informal proof provided in the core of the paper should be complemented by formal proofs provided in appendix or supplemental material.
        \item Theorems and Lemmas that the proof relies upon should be properly referenced. 
    \end{itemize}

    \item {\bf Experimental result reproducibility}
    \item[] Question: Does the paper fully disclose all the information needed to reproduce the main experimental results of the paper to the extent that it affects the main claims and/or conclusions of the paper (regardless of whether the code and data are provided or not)?
    \item[] Answer: \answerYes{}
    \item[] Justification: The experiments are conducted on open-source models and three datasets—two of which are publicly available, and one is a synthetic dataset, with detailed construction described in Appendix~\ref{append:templates}. The full experimental configurations are provided in Section~\ref{sec:exp-setting}.
    \item[] Guidelines:
    \begin{itemize}
        \item The answer NA means that the paper does not include experiments.
        \item If the paper includes experiments, a No answer to this question will not be perceived well by the reviewers: Making the paper reproducible is important, regardless of whether the code and data are provided or not.
        \item If the contribution is a dataset and/or model, the authors should describe the steps taken to make their results reproducible or verifiable. 
        \item Depending on the contribution, reproducibility can be accomplished in various ways. For example, if the contribution is a novel architecture, describing the architecture fully might suffice, or if the contribution is a specific model and empirical evaluation, it may be necessary to either make it possible for others to replicate the model with the same dataset, or provide access to the model. In general. releasing code and data is often one good way to accomplish this, but reproducibility can also be provided via detailed instructions for how to replicate the results, access to a hosted model (e.g., in the case of a large language model), releasing of a model checkpoint, or other means that are appropriate to the research performed.
        \item While NeurIPS does not require releasing code, the conference does require all submissions to provide some reasonable avenue for reproducibility, which may depend on the nature of the contribution. For example
        \begin{enumerate}
            \item If the contribution is primarily a new algorithm, the paper should make it clear how to reproduce that algorithm.
            \item If the contribution is primarily a new model architecture, the paper should describe the architecture clearly and fully.
            \item If the contribution is a new model (e.g., a large language model), then there should either be a way to access this model for reproducing the results or a way to reproduce the model (e.g., with an open-source dataset or instructions for how to construct the dataset).
            \item We recognize that reproducibility may be tricky in some cases, in which case authors are welcome to describe the particular way they provide for reproducibility. In the case of closed-source models, it may be that access to the model is limited in some way (e.g., to registered users), but it should be possible for other researchers to have some path to reproducing or verifying the results.
        \end{enumerate}
    \end{itemize}

\item {\bf Open access to data and code}
    \item[] Question: Does the paper provide open access to the data and code, with sufficient instructions to faithfully reproduce the main experimental results, as described in supplemental material?
    \item[] Answer: \answerNo{}
    \item[] Justification: We do not have the time to refactor the code, which is of poor readability. We promise to open-source the code to reproduce the experimental results on GitHub once accepted.
    \item[] Guidelines:
    \begin{itemize}
        \item The answer NA means that paper does not include experiments requiring code.
        \item Please see the NeurIPS code and data submission guidelines (\url{https://nips.cc/public/guides/CodeSubmissionPolicy}) for more details.
        \item While we encourage the release of code and data, we understand that this might not be possible, so “No” is an acceptable answer. Papers cannot be rejected simply for not including code, unless this is central to the contribution (e.g., for a new open-source benchmark).
        \item The instructions should contain the exact command and environment needed to run to reproduce the results. See the NeurIPS code and data submission guidelines (\url{https://nips.cc/public/guides/CodeSubmissionPolicy}) for more details.
        \item The authors should provide instructions on data access and preparation, including how to access the raw data, preprocessed data, intermediate data, and generated data, etc.
        \item The authors should provide scripts to reproduce all experimental results for the new proposed method and baselines. If only a subset of experiments are reproducible, they should state which ones are omitted from the script and why.
        \item At submission time, to preserve anonymity, the authors should release anonymized versions (if applicable).
        \item Providing as much information as possible in supplemental material (appended to the paper) is recommended, but including URLs to data and code is permitted.
    \end{itemize}

\item {\bf Experimental setting/details}
    \item[] Question: Does the paper specify all the training and test details (e.g., data splits, hyperparameters, how they were chosen, type of optimizer, etc.) necessary to understand the results?
    \item[] Answer: \answerYes{}
    \item[] Justification: We follow the standard method of prior work as we have mentioned in Section \ref{sec:exp-setting}.
    \item[] Guidelines:
    \begin{itemize}
        \item The answer NA means that the paper does not include experiments.
        \item The experimental setting should be presented in the core of the paper to a level of detail that is necessary to appreciate the results and make sense of them.
        \item The full details can be provided either with the code, in appendix, or as supplemental material.
    \end{itemize}

\item {\bf Experiment statistical significance}
    \item[] Question: Does the paper report error bars suitably and correctly defined or other appropriate information about the statistical significance of the experiments?
    \item[] Answer: \answerYes{}
    \item[] Justification: We conduct the experiments many times and report average results. In Figure~\ref{fig:cs-strategy}, we show the one standard deviation as a shaded region.
    \item[] Guidelines:
    \begin{itemize}
        \item The answer NA means that the paper does not include experiments.
        \item The authors should answer "Yes" if the results are accompanied by error bars, confidence intervals, or statistical significance tests, at least for the experiments that support the main claims of the paper.
        \item The factors of variability that the error bars are capturing should be clearly stated (for example, train/test split, initialization, random drawing of some parameter, or overall run with given experimental conditions).
        \item The method for calculating the error bars should be explained (closed form formula, call to a library function, bootstrap, etc.)
        \item The assumptions made should be given (e.g., Normally distributed errors).
        \item It should be clear whether the error bar is the standard deviation or the standard error of the mean.
        \item It is OK to report 1-sigma error bars, but one should state it. The authors should preferably report a 2-sigma error bar than state that they have a 96\% CI, if the hypothesis of Normality of errors is not verified.
        \item For asymmetric distributions, the authors should be careful not to show in tables or figures symmetric error bars that would yield results that are out of range (e.g. negative error rates).
        \item If error bars are reported in tables or plots, The authors should explain in the text how they were calculated and reference the corresponding figures or tables in the text.
    \end{itemize}

\item {\bf Experiments compute resources}
    \item[] Question: For each experiment, does the paper provide sufficient information on the computer resources (type of compute workers, memory, time of execution) needed to reproduce the experiments?
    \item[] Answer: \answerYes{}
    \item[] Justification: We mention that we run the experiments on 1 NVIDIA V100 GPU. The specific configurations are included in the Section \ref{sec:exp-setting}.
    \item[] Guidelines:
    \begin{itemize}
        \item The answer NA means that the paper does not include experiments.
        \item The paper should indicate the type of compute workers CPU or GPU, internal cluster, or cloud provider, including relevant memory and storage.
        \item The paper should provide the amount of compute required for each of the individual experimental runs as well as estimate the total compute. 
        \item The paper should disclose whether the full research project required more compute than the experiments reported in the paper (e.g., preliminary or failed experiments that didn't make it into the paper). 
    \end{itemize}
    
\item {\bf Code of ethics}
    \item[] Question: Does the research conducted in the paper conform, in every respect, with the NeurIPS Code of Ethics \url{https://neurips.cc/public/EthicsGuidelines}?
    \item[] Answer: \answerYes{}
    \item[] Justification: We follow the NeurIPS Code of Ethics.
    \item[] Guidelines:
    \begin{itemize}
        \item The answer NA means that the authors have not reviewed the NeurIPS Code of Ethics.
        \item If the authors answer No, they should explain the special circumstances that require a deviation from the Code of Ethics.
        \item The authors should make sure to preserve anonymity (e.g., if there is a special consideration due to laws or regulations in their jurisdiction).
    \end{itemize}

\item {\bf Broader impacts}
    \item[] Question: Does the paper discuss both potential positive societal impacts and negative societal impacts of the work performed?
    \item[] Answer: \answerYes{} 
    \item[] Justification: We discuss the potential positive societal impacts of using \name for protecting the prompt privacy in the Introduction Section~\ref{sec:intro}. More detailed elaboration is provided in Appendix~\ref{append:broad-impact}.
    \item[] Guidelines:
    \begin{itemize}
        \item The answer NA means that there is no societal impact of the work performed.
        \item If the authors answer NA or No, they should explain why their work has no societal impact or why the paper does not address societal impact.
        \item Examples of negative societal impacts include potential malicious or unintended uses (e.g., disinformation, generating fake profiles, surveillance), fairness considerations (e.g., deployment of technologies that could make decisions that unfairly impact specific groups), privacy considerations, and security considerations.
        \item The conference expects that many papers will be foundational research and not tied to particular applications, let alone deployments. However, if there is a direct path to any negative applications, the authors should point it out. For example, it is legitimate to point out that an improvement in the quality of generative models could be used to generate deepfakes for disinformation. On the other hand, it is not needed to point out that a generic algorithm for optimizing neural networks could enable people to train models that generate Deepfakes faster.
        \item The authors should consider possible harms that could arise when the technology is being used as intended and functioning correctly, harms that could arise when the technology is being used as intended but gives incorrect results, and harms following from (intentional or unintentional) misuse of the technology.
        \item If there are negative societal impacts, the authors could also discuss possible mitigation strategies (e.g., gated release of models, providing defenses in addition to attacks, mechanisms for monitoring misuse, mechanisms to monitor how a system learns from feedback over time, improving the efficiency and accessibility of ML).
    \end{itemize}
    
\item {\bf Safeguards}
    \item[] Question: Does the paper describe safeguards that have been put in place for responsible release of data or models that have a high risk for misuse (e.g., pretrained language models, image generators, or scraped datasets)?
    \item[] Answer: \answerNA{}
    \item[] Justification: The answer NA means that the paper poses no such risks.
    \item[] Guidelines:
    \begin{itemize}
        \item The answer NA means that the paper poses no such risks.
        \item Released models that have a high risk for misuse or dual-use should be released with necessary safeguards to allow for controlled use of the model, for example by requiring that users adhere to usage guidelines or restrictions to access the model or implementing safety filters. 
        \item Datasets that have been scraped from the Internet could pose safety risks. The authors should describe how they avoided releasing unsafe images.
        \item We recognize that providing effective safeguards is challenging, and many papers do not require this, but we encourage authors to take this into account and make a best faith effort.
    \end{itemize}

\item {\bf Licenses for existing assets}
    \item[] Question: Are the creators or original owners of assets (e.g., code, data, models), used in the paper, properly credited and are the license and terms of use explicitly mentioned and properly respected?
    \item[] Answer: \answerYes{} % Replace by \answerYes{}, \answerNo{}, or \answerNA{}.
    \item[] Justification: We cite the original paper that produced the code, dataset and models.
    \item[] Guidelines: 
    \begin{itemize}
        \item The answer NA means that the paper does not use existing assets.
        \item The authors should cite the original paper that produced the code package or dataset.
        \item The authors should state which version of the asset is used and, if possible, include a URL.
        \item The name of the license (e.g., CC-BY 4.0) should be included for each asset.
        \item For scraped data from a particular source (e.g., website), the copyright and terms of service of that source should be provided.
        \item If assets are released, the license, copyright information, and terms of use in the package should be provided. For popular datasets, \url{paperswithcode.com/datasets} has curated licenses for some datasets. Their licensing guide can help determine the license of a dataset.
        \item For existing datasets that are re-packaged, both the original license and the license of the derived asset (if it has changed) should be provided.
        \item If this information is not available online, the authors are encouraged to reach out to the asset's creators.
    \end{itemize}

\item {\bf New assets}
    \item[] Question: Are new assets introduced in the paper well documented and is the documentation provided alongside the assets?
    \item[] Answer: \answerNA{} % Replace by \answerYes{}, \answerNo{}, or \answerNA{}.
    \item[] Justification: This paper does not release new assets.
    \item[] Guidelines:
    \begin{itemize}
        \item The answer NA means that the paper does not release new assets.
        \item Researchers should communicate the details of the dataset/code/model as part of their submissions via structured templates. This includes details about training, license, limitations, etc. 
        \item The paper should discuss whether and how consent was obtained from people whose asset is used.
        \item At submission time, remember to anonymize your assets (if applicable). You can either create an anonymized URL or include an anonymized zip file.
    \end{itemize}

\item {\bf Crowdsourcing and research with human subjects}
    \item[] Question: For crowdsourcing experiments and research with human subjects, does the paper include the full text of instructions given to participants and screenshots, if applicable, as well as details about compensation (if any)? 
    \item[] Answer: \answerNA{} % Replace by \answerYes{}, \answerNo{}, or \answerNA{}.
    \item[] Justification: This paper does not involve crowdsourcing nor research with human subjects.
    \item[] Guidelines:
    \begin{itemize}
        \item The answer NA means that the paper does not involve crowdsourcing nor research with human subjects.
        \item Including this information in the supplemental material is fine, but if the main contribution of the paper involves human subjects, then as much detail as possible should be included in the main paper. 
        \item According to the NeurIPS Code of Ethics, workers involved in data collection, curation, or other labor should be paid at least the minimum wage in the country of the data collector. 
    \end{itemize}

\item {\bf Institutional review board (IRB) approvals or equivalent for research with human subjects}
    \item[] Question: Does the paper describe potential risks incurred by study participants, whether such risks were disclosed to the subjects, and whether Institutional Review Board (IRB) approvals (or an equivalent approval/review based on the requirements of your country or institution) were obtained?
    \item[] Answer: \answerNA{} % Replace by \answerYes{}, \answerNo{}, or \answerNA{}.
    \item[] Justification: This paper does not involve crowdsourcing nor research with human subjects.
    \item[] Guidelines:
    \begin{itemize}
        \item The answer NA means that the paper does not involve crowdsourcing nor research with human subjects.
        \item Depending on the country in which research is conducted, IRB approval (or equivalent) may be required for any human subjects research. If you obtained IRB approval, you should clearly state this in the paper. 
        \item We recognize that the procedures for this may vary significantly between institutions and locations, and we expect authors to adhere to the NeurIPS Code of Ethics and the guidelines for their institution. 
        \item For initial submissions, do not include any information that would break anonymity (if applicable), such as the institution conducting the review.
    \end{itemize}

\item {\bf Declaration of LLM usage}
    \item[] Question: Does the paper describe the usage of LLMs if it is an important, original, or non-standard component of the core methods in this research? Note that if the LLM is used only for writing, editing, or formatting purposes and does not impact the core methodology, scientific rigorousness, or originality of the research, declaration is not required.
    \item[] Answer: \answerNA{} % Replace by \answerYes{}, \answerNo{}, or \answerNA{}.
    \item[] Justification: We only use the LLM for paper writing, editing, or formatting purposes.
    \item[] Guidelines:
    \begin{itemize}
        \item The answer NA means that the core method development in this research does not involve LLMs as any important, original, or non-standard components.
        \item Please refer to our LLM policy (\url{https://neurips.cc/Conferences/2025/LLM}) for what should or should not be described.
    \end{itemize}
    
\end{enumerate}

%% file: tex/appendix.tex
\section{Broad Impacts}\label{append:broad-impact}
This paper presents work whose goal is to advance the field of Machine Learning. Our work aims to address critical privacy concerns surrounding user prompts in widely adopted stable diffusion based text-to-image generation services.
To safeguard sensitive user information during inference, we propose an oblivious cloud-device hybrid image generation scheme, acting as a plug-and-play safeguard to ML inference services, to provide rigorous prompt privacy. We also acknowledge the redundant computational overhead introduced by generating intermediate latents for all candidate prompts in oblivious generation. Beyond our proposed acceleration strategies, further improvements could be achieved by constraining the sampling space of candidate prompts. Ultimately, this reflects a fundamental trade-off between privacy and efficiency.
These advancements have the potential to make private LLM inference more practical and scalable for real-world applications.

\section{Implementation Details}\label{append:exp-details}

\subsection{Oblivious Cloud-Device Hybrid Generation Algorithm}\label{append:hybrid-algo}
The detailed design is presented in Algorithm~\ref{algo:hybrid-oblivious}.
Note that in line 2, \texttt{all\_combination} refers to enumerating all possible combinations by traversing every value of each attribute $F_i \in \mathcal{F}_{\text{occur}}$, where $F_i = \{f_1, \dots, f_q\}$.

\begin{algorithm}
\caption{Oblivious Hybrid Generation}\label{algo:hybrid-oblivious}

    \KwIn{User prompt $p = \{w_1, ...,w_i,..., w_n\}$, $w_i$ corresponds to some sensitive attribute $F_i \in \mathcal{F}=\{F_1,...,F_m\}$, with a value space of size $q$ as $F_i = \{f_1,...,f_q\}$, a client-side small diffusion model $\mathcal{M}_C$ and a small language model $\mathcal{M}_{LM}$, a server-side large model $\mathcal{M}_S$, denoising steps $T = \{t_{T-1},...,t_0\}$, switch point $k$, random latent $z_T\sim \mathcal{N}(0, \mathbf{I})$, VAE decoder $\mathcal{D}(\cdot)$}
    \KwOut{Generated image $y$}

    \tcc{\textbf{Oblivious Transform}: construct candidate prompts based on $p$}
    
    $\mathcal{F}_{occur} = \mathcal{M}_{LM}(p)$, with $F_i \in \mathcal{F}_{occur} = \{f_1,...,f_q\}$ \\

    $\mathcal{P} \leftarrow \texttt{all\_combination}(p, \mathcal{F}_{occur})$ \\

    $\hat{z}_T = \mathrm{repeat}(z_T, N)$, with $N= |\mathcal{P}|$ \\

    \tcc{Run initial $k$ steps on cloud-side}

    \For{$t \in T[:k]$}{
        $\hat{z}_{t-1} = \mathcal{M}_S(\hat{z}_t, \mathcal{P}, t)$
    }

    \tcc{\textbf{Extraction}:$\hat{z}_{T-k}$ are sent to device, who chooses the actual latent $z_{T-k}$}
    
    $z_{T-k} = \hat{z}_{T-k,j}, \text{s.t. } p_j \in \mathcal{P} \ \text{and}\ p_j = p$ 
    
    \tcc{Run remaining $T-k$ steps on device-side}

    \For{$t \in T[k:]$}{
        $z_{t-1} = \mathcal{M}_C(z_t, p, t)$
    }

    $y = \mathcal{D}(z_0)$
    
    \KwRet{y}
\end{algorithm}

\subsection{Batch Reuse Algorithm}\label{append:batch-reuse-algo}
In this section, we provide the batch-reuse in attention modules in Algorithm~\ref{alg:batch_reuse}. 
\begin{algorithm}[ht]
\caption{Batch-reused Attention Module}
\label{alg:batch_reuse}
\KwIn{Hidden states $\mathcal{Q} = \{q_i\}_{i=1}^N$, Encoder hidden states $\mathcal{K} = \{k_i\}_{i=1}^N, \mathcal{V} = \{v_i\}_{i=1}^N$ for $N$ candidate prompts, Pivot sample index $i^*$ for pivot prompt $p$}
\KwOut{Attention outputs $O = \{o_i\}_{i=1}^N$}

\tcc{Compute query $q^*$, key $k^*$ for pivot prompt $p$ (e.g., $i^*=0$)}

$q^* = \texttt{to\_q}(\mathcal{Q}[i^*])$ \\
$k^* = \texttt{to\_k}(\mathcal{K}[i^*])$ \\

\tcc{Compute attention map for pivot prompt}

$m^* = \texttt{get\_attention\_map}(q^*, k^*)$ \\

\tcc{Compute value $v$ for all candidate prompts}
$V = \texttt{to\_v}(\mathcal{V})$ \\

\tcc{Compute attention outputs for all candidate prompts}

$M = \texttt{broadcast}(m^*, N)$ \\
$O = M \cdot V$

\Return{$O$}
\end{algorithm}

\subsection{Candidate Prompt Dataset Construction}\label{append:templates}
\captionsetup{labelfont=bf}
\lstdefinelanguage{yaml}{
    keywords={true,false,null,y,n},
    keywordstyle=\color{red}\bfseries,
    basicstyle=\ttfamily\small,
    sensitive=false,
    comment=[l]{\#},
    morecomment=[s]{/*}{*/},
    commentstyle=\color{gray}\ttfamily,
    stringstyle=\color{orange}\ttfamily,
    moredelim=[l][\color{green}]{---},
    moredelim=[l][\color{green}]{...},
    moredelim=[l][\color{cyan}]{-}
}
\lstset{ %
  language=yaml,                % use the YAML language
  basicstyle=\ttfamily\scriptsize, % the size of the fonts that are used for the code
  numbers=none,                   % where to put the line-numbers
  numberstyle=\tiny\color{gray},  % the style that is used for the line-numbers
  stepnumber=1,                   % the step between two line-numbers. If it's 1, each line will be numbered
  numbersep=5pt,                  % how far the line-numbers are from the code
  backgroundcolor=\color{yellow!10},  % choose the background color. You must add \usepackage{color}
  showspaces=false,               % show spaces adding particular underscores
  showstringspaces=false,         % underline spaces within strings
  showtabs=false,                 % show tabs within strings adding particular underscores
  frame=none,                   % adds a frame around the code
  rulecolor=\color{black},        % if not set, the frame-color may be changed on line-breaks within not-black text (e.g. comments (green here))
  tabsize=2,                      % sets default tabsize to 2 spaces
  captionpos=b,                   % sets the caption-position to bottom
  breaklines=true,                % sets automatic line breaking
  breakatwhitespace=false,        % sets if automatic breaks should only happen at whitespace
  title=\lstname,                 % show the filename of files included with \lstinputlisting; also try caption instead of title
  keywordstyle=\bfseries\color{red},      % keyword style
  commentstyle=\color{dkgreen},   % comment style
  stringstyle=\color{mauve},      % string literal style
  escapeinside={\%*}{*)},         % if you want to add LaTeX within your code
  morekeywords={tags, detection}            % if you want to add more keywords to the set
}
\begin{lstlisting}[caption={Prompt templates and sensitive attributes taxonomy.}, label=lst:sigmarule]
template:
    - "headshots portrait with a $age $ethnicity $gender covered in religious tattoos."
    - "$age $ethnicity $gender in hat Fashion portrait photo"
    - "Smiling $age $ethnicity $gender sitting on flower field, Outdoor portrait photo"
    - "$age red haired $gender $ethnicity urban portrait photo"
    - "Faceshot Portrait of pretty $age $ethnicity $gender wearing a high neck sweater"
    - Closeup portrait photo of a $age $ethnicity $gender, wearing a rugged leather jacket, with a five o'clock shadow and prominent laugh lines around his eyes, captured in soft, golden hour lighting."
    - "RAW photo, (closeup:1.2), portrait of a $age $ethnicity $gender, wearing minimal makeup, showcasing the freckles, with a serene expression in a lush botanical garden, illuminated by gentle dappled sunlight."
    - "High-quality, face portrait photo of a $age $ethnicity $gender, wearing glasses, revealing the fine lines and character on the forehead."
    - "B&W photo of a $age $ethnicity $gender, shot from the side, highlighting elegant profile and the delicate lines etched across cheeks."
    - "High-quality, closeup portrait photo of a $age $ethnicity $gender, wearing traditional clothing."
age:
    - young
    - middle-aged
    - old  
gender:
    - male
    - female
ethnicity:
    - caucasian
    - african
    - asian
    - indian
    - european
\end{lstlisting}

We list the 10 templates used in constructing a small prompt dataset, tailored for candidate prompts, with replacement on sensitive attributes age, gender and ethnicity.
We consider \textit{age} $\in$ \{young, middle-aged, old\}, \textit{gender} $\in$ \{male, female\} and \textit{ethnicity} $\in$ \{caucasian, african, asian, indian, european\}.

\section{Security Analysis}\label{append:security}
\begin{proof}[Proof of Theorem~\ref{theorem:oblivious}]
    Consider a user prompt $p$. We obtain its candidate prompts as  $\mathcal{P} = \texttt{candidate\_prompt}(p)$ by traversing the entire value space for each sensitive attribute in $p$. 
    Then, we randomly select another sensitive prompt $p'\in \mathcal{P}$. 
    Since we traverse the sensitive attributes to get $\mathcal{P}$, we have $\mathcal{P} = \texttt{candidate\_prompt}(p) = \texttt{candidate\_prompt}(p')$. That is,
    \begin{equation*}
        \textbf{View}(p) = \textbf{View}(p'), \forall p' \in \mathcal{P}    
    \end{equation*}

    We thus have, for any two sensitive prompts $p, p'\in \mathcal{P}$, the server's observable view is identically distributed. That is, no efficient adversary can guess the correct $p$ with significantly better probability than random guessing ($1/N$) as:
    \begin{equation*}
        |\mathrm{Pr}[A(\mathcal{P} = p^*)] - \frac{1}{N}| \leq \lambda
    \end{equation*}
    where $\lambda$ is negligible.
    The proof shows that the server learns nothing about the true input $p$ beyond what is leaked by the candidate prompts (i.e., the value space for sensitive attributes) and those non-sensitive tokens, which are independent of the specific sensitive tokens.
\end{proof}

\section{Additional Experiments}\label{append:additional-exps}

\subsection{Self-Attention Visualization}\label{append:self-attn-visualization}
\begin{figure}[ht]
    \centering
    \includegraphics[width=0.8\linewidth]{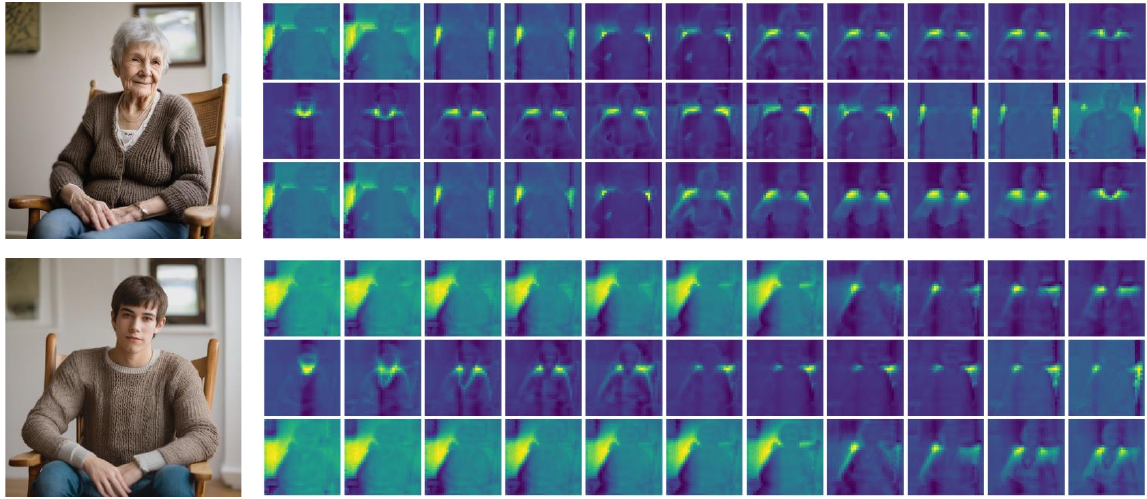}
    \caption{The top components obtained using SVD of self-attention maps for two candidate prompts.}
    \label{fig:self-attn-vis}
\end{figure}
Figure~\ref{fig:self-attn-vis} illustrates the self-attention maps for two candidate prompts by varying \textit{gender} and \textit{age} attributes. Specifically, from ``An \underline{elderly woman} with short hair, wearing a knitted sweater, sitting in a rocking chair.'' to ``A \underline{young man} with short hair, wearing a knitted sweater, sitting in a rocking chair.''. We capture the self-attention maps in middle layers and run SVD to obtain the top components.

\subsection{Model Statistics}~\label{append:model-statistics}
Table~\ref{tab:model-details} presents the quantitative statistics for standalone cloud and device models, including utility scores, parameter size and FLOPs consumption. We adopt 25-step DPM scheduler for all these models by default. For LCM-SDXL, we use 8-step LCM scheduler instead. The FLOPs and latency are measured for a batch size of 4. We run each experiment ten times and report the average results.

\begin{table}[ht]
    \centering
    \caption{Model statistics. By default, we use 25-step DPM scheduler. $B=4$.}
    \scalebox{0.85}{
    \begin{tabular}{@{}lcccccc@{}}
\toprule
\multicolumn{1}{c|}{\textbf{Generation Method}} & \multicolumn{1}{c|}{\textbf{FID $\downarrow$}} & \multicolumn{1}{c|}{\textbf{IS $\uparrow$}} & \multicolumn{1}{c|}{\textbf{CLIP $\uparrow$}} & \multicolumn{1}{c|}{\textbf{\#Params (M)}} & \multicolumn{1}{c|}{\textbf{FLOPs (T)}} & \textbf{Latency (s)} \\ \midrule
\multicolumn{7}{c}{MS-COCO}                                                                                                                                                                                                                                                                                  \\ \midrule
\multicolumn{1}{l|}{SD-v1.4}                    & \multicolumn{1}{c|}{13.86}                     & \multicolumn{1}{c|}{37.75}                  & \multicolumn{1}{c|}{0.3015}                   & \multicolumn{1}{c|}{859.40}                & \multicolumn{1}{c|}{74.10}              & 5.01                 \\
\multicolumn{1}{l|}{BK-SDM-small}               & \multicolumn{1}{c|}{18.30}                     & \multicolumn{1}{c|}{31.73}                  & \multicolumn{1}{c|}{0.2710}                   & \multicolumn{1}{c|}{482.28}                & \multicolumn{1}{c|}{43.60}              & 3.01                 \\
\multicolumn{1}{l|}{BK-SDM-tiny}                & \multicolumn{1}{c|}{18.30}                     & \multicolumn{1}{c|}{29.94}                  & \multicolumn{1}{c|}{0.2681}                   & \multicolumn{1}{c|}{323.34}                & \multicolumn{1}{c|}{41.00}              & 2.87                 \\ \midrule
\multicolumn{1}{l|}{Realistic Vision v4.0}      & \multicolumn{1}{c|}{16.21}                     & \multicolumn{1}{c|}{37.39}                  & \multicolumn{1}{c|}{0.3033}                   & \multicolumn{1}{c|}{859.40}                & \multicolumn{1}{c|}{74.10}              & 4.42                 \\
\multicolumn{1}{l|}{small-sd}                   & \multicolumn{1}{c|}{14.59}                     & \multicolumn{1}{c|}{35.06}                  & \multicolumn{1}{c|}{0.3046}                   & \multicolumn{1}{c|}{579.30}                & \multicolumn{1}{c|}{44.80}              & 2.96                 \\ \midrule
\multicolumn{7}{c}{MJHQ}                                                                                                                                                                                                                                                                                     \\ \midrule
\multicolumn{1}{l|}{SDXL}                       & \multicolumn{1}{c|}{30.67}                     & \multicolumn{1}{c|}{26.31}                  & \multicolumn{1}{c|}{0.3464}                   & \multicolumn{1}{c|}{2562.13}               & \multicolumn{1}{c|}{637.40}             & 29.33                \\
\multicolumn{1}{l|}{koala-700m}                 & \multicolumn{1}{c|}{34.11}                     & \multicolumn{1}{c|}{22.06}                  & \multicolumn{1}{c|}{0.3263}                   & \multicolumn{1}{c|}{2562.13}               & \multicolumn{1}{c|}{235.40}             & 12.00                \\
\multicolumn{1}{l|}{LCM-SDXL (8-step)}          & \multicolumn{1}{c|}{33.25}                     & \multicolumn{1}{c|}{28.22}                  & \multicolumn{1}{c|}{0.3296}                   & \multicolumn{1}{c|}{777.53}                & \multicolumn{1}{c|}{203.97}             & 8.35                 \\ \bottomrule
\end{tabular}
}
    \label{tab:model-details}
\end{table}

\subsection{Visualization of Hybrid Generation}\label{append:vis-hybrid-gen}

\begin{figure}[ht]
    \centering
    \includegraphics[width=0.95\linewidth]{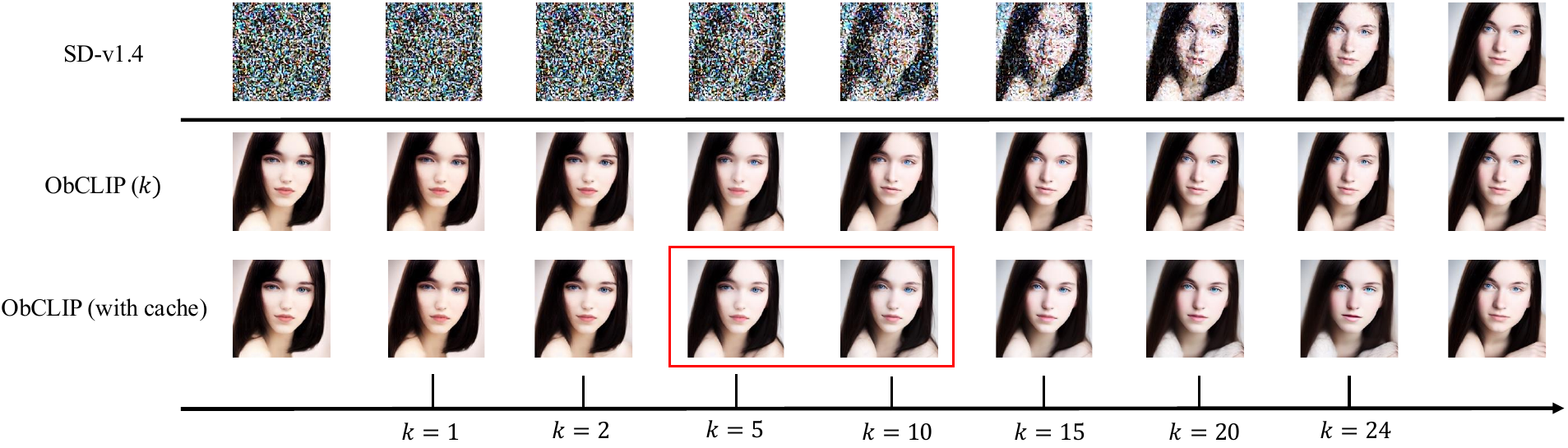}
    \caption{Visualization of SD-v1.4 + BK-SDM-small}
    \label{fig:vis-sd1.4}
\end{figure}

\begin{figure}[ht]
    \centering
    \includegraphics[width=0.95\linewidth]{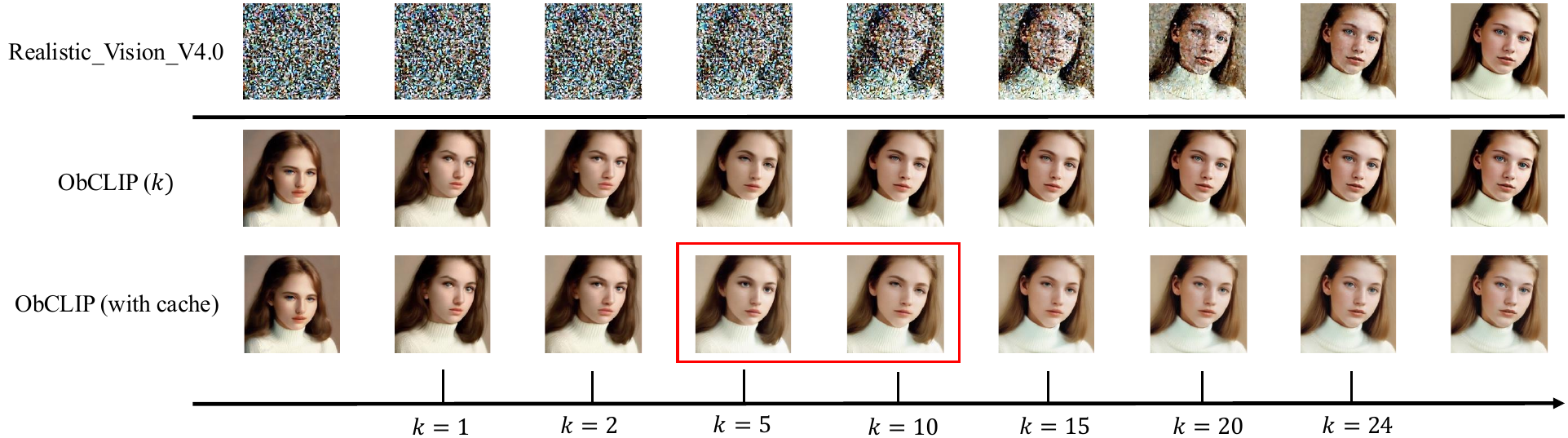}
    \caption{Visualization of Realisti-Vision-V4.0 + small-sd}
    \label{fig:vis-realistic}
\end{figure}

In this section, we provide the visualization of generated images for large server model (top), and \name with/without server-side acceleration (middle and bottom) in Figure~\ref{fig:vis-sd1.4} and Figure~\ref{fig:vis-realistic}. The used prompt is ``\textit{Faceshot Portrait of pretty young (18-year-old) female Caucasian wearing a high neck sweater}''. 
We mark the two images with an optimal balance between image quality and server computation cost, i.e., when $k\in \{5, 10\}$. The better semantic from large server model are well preserved with minimal server denoising cost.

\subsection{3-Attribute Oblivious Hybrid Generation}\label{append:3-attribute}

The detailed utility scores for 3-attribute oblivious hybrid image generation are presented in Table~\ref{tab:3-attribute}. Compared to the large server-side model, there is only a marginal drop in image quality and text-image alignment, while achieving significantly better utility than the on-device small model.
\begin{table}[ht]
    \centering
    \caption{Evaluation results for 3-attribute oblivious generation.}
    \scalebox{1.0}{
    \begin{tabular}{l|c|c|c}
\toprule
\textbf{Generation Method}   & \textbf{FID $\downarrow$} & \textbf{IS $\uparrow$} & \textbf{CLIP $\uparrow$} \\ \midrule
Realistic                    & 111.87                    & 4.78                   & 0.3322                   \\ \midrule
small-sd                     & 115.96                    & 5.02                   & 0.3034                   \\ \midrule
Vanilla OG (Realistic)       & 111.87                    & 4.78                   & 0.3322                   \\ \midrule
OG (k = 10)                  & 112.75                    & 5.04                   & 0.3214                   \\
OG (k = 10) w cache          & 113.31                    & 5.03                   & 0.3171                   \\
\rowcolor[HTML]{C0C0C0} 
OG (k = 10) w cache + reuse  & 110.22                    & 4.62                   & 0.3138                   \\ \midrule
OG (k = 5)                   & 113.57                    & 5.02                   & 0.3108                   \\
OG (k = 5) (w cache)         & 114.07                    & 5.04                   & 0.3083                   \\
\rowcolor[HTML]{9B9B9B} 
OG (k = 5) (w cache) + reuse & 113.30                    & 4.66                   & 0.3077                   \\ \bottomrule
\end{tabular}
}
    \label{tab:3-attribute}
\end{table}

\subsection{Comprehensive Efficiency Evaluation}\label{append:efficiency}

We here present a more comprehensive efficiency evaluation across different candidate prompt size $N$ to validate the effectiveness of \name. As shown in Table~\ref{tab:efficiency}, in most cases, \name achieves server-side efficiency comparable to that of server-only image generation on both SD-v1.4 and SDXL models. When accounting for device-side computation, the total latency remains of the same order of magnitude as the baseline. These results demonstrate the superior performance of \name, which offers rigorous privacy with only a slight increase in overall computation cost.

\begin{table}[ht]
    \centering
    \caption{Comprehensive efficiency evaluation.}
    \scalebox{0.8}{
    \begin{tabular}{@{}llcccccccc@{}}
\toprule
\multicolumn{2}{c|}{}                                                                                                        & \multicolumn{4}{c|}{\textbf{FLOPs}}                                                                                                                                                                                  & \multicolumn{4}{c}{\textbf{Latency (s)}}                                                                                                                                                 \\ \cmidrule(l){3-10} 
\multicolumn{2}{c|}{}                                                                                                        & \multicolumn{2}{c|}{\textbf{$k=5$}}                                                                      & \multicolumn{2}{c|}{\textbf{$k=10$}}                                                                      & \multicolumn{2}{c|}{\textbf{$k=5$}}                                                                   & \multicolumn{2}{c}{\textbf{$k=10$}}                                              \\ \cmidrule(l){3-10} 
\multicolumn{2}{c|}{\multirow{-3}{*}{\textbf{Generation Method}}}                                                            & \multicolumn{1}{c|}{\textbf{$N=4$}}                & \multicolumn{1}{c|}{\textbf{$N=6$}}                 & \multicolumn{1}{c|}{\textbf{$N=4$}}                 & \multicolumn{1}{c|}{\textbf{$N=6$}}                 & \multicolumn{1}{c|}{\textbf{$N=4$}}               & \multicolumn{1}{c|}{\textbf{$N=6$}}               & \multicolumn{1}{c|}{\textbf{$N=4$}}               & \textbf{$N=6$}               \\ \midrule
\multicolumn{10}{c}{SD (SD-v1.4 + BK-SDM-small)}                                                                                                                                                                                                                                                                                                                                                                                                                                                                                               \\ \midrule
\multicolumn{1}{l|}{}                              & \multicolumn{1}{l|}{SD-v1.4}                                            & \multicolumn{4}{c|}{18.53}                                                                                                                                                                                           & \multicolumn{4}{c}{1.28}                                                                                                                                                                 \\ \cmidrule(l){3-10} 
\multicolumn{1}{l|}{\multirow{-2}{*}{Non-Private}} & \multicolumn{1}{l|}{Hybrid SD}                                          & \multicolumn{2}{c|}{3.71}                                                                                & \multicolumn{2}{c|}{7.41}                                                                                 & \multicolumn{2}{c|}{0.30}                                                                             & \multicolumn{2}{c}{0.51}                                                         \\ \midrule
\multicolumn{1}{l|}{}                              & \multicolumn{1}{l|}{HE-Diffusion}                                       & \multicolumn{4}{c|}{-}                                                                                                                                                                                               & \multicolumn{4}{c}{$> 106$}                                                                                                                                                              \\ \cmidrule(l){2-10} 
\multicolumn{1}{l|}{}                              & \multicolumn{1}{l|}{Vanilla OG}                                         & \multicolumn{1}{c|}{74.10}                         & \multicolumn{1}{c|}{111.15}                         & \multicolumn{1}{c|}{74.10}                          & \multicolumn{1}{c|}{111.15}                         & \multicolumn{1}{c|}{5.01}                         & \multicolumn{1}{c|}{7.47}                         & \multicolumn{1}{c|}{5.01}                         & 7.47                         \\
\multicolumn{1}{l|}{}                              & \multicolumn{1}{l|}{$\name$}                                            & \multicolumn{1}{c|}{14.82}                         & \multicolumn{1}{c|}{22.23}                          & \multicolumn{1}{c|}{29.64}                          & \multicolumn{1}{c|}{44.46}                          & \multicolumn{1}{c|}{1.11}                         & \multicolumn{1}{c|}{1.48}                         & \multicolumn{1}{c|}{1.93}                         & 2.90                         \\
\multicolumn{1}{l|}{}                              & \multicolumn{1}{l|}{$\enspace\textbf{+}$ cache}                         & \multicolumn{1}{c|}{12.25}                         & \multicolumn{1}{c|}{18.38}                          & \multicolumn{1}{c|}{23.36}                          & \multicolumn{1}{c|}{35.04}                          & \multicolumn{1}{c|}{0.67}                         & \multicolumn{1}{c|}{1.12}                         & \multicolumn{1}{c|}{1.33}                         & 1.85                         \\
\multicolumn{1}{l|}{}                              & \multicolumn{1}{l|}{\cellcolor[HTML]{9B9B9B}$\enspace\textbf{+}$ reuse} & \multicolumn{1}{c|}{\cellcolor[HTML]{9B9B9B}11.13} & \multicolumn{1}{c|}{\cellcolor[HTML]{9B9B9B}16.64}  & \multicolumn{1}{c|}{\cellcolor[HTML]{9B9B9B}21.86}  & \multicolumn{1}{c|}{\cellcolor[HTML]{9B9B9B}32.72}  & \multicolumn{1}{c|}{\cellcolor[HTML]{9B9B9B}0.61} & \multicolumn{1}{c|}{\cellcolor[HTML]{9B9B9B}0.98} & \multicolumn{1}{c|}{\cellcolor[HTML]{9B9B9B}1.12} & \cellcolor[HTML]{9B9B9B}1.55 \\ \cmidrule(l){2-10} 
\multicolumn{1}{l|}{\multirow{-6}{*}{Private}}     & \multicolumn{1}{l|}{Total ($\enspace\textbf{+}$ device)}                & \multicolumn{1}{c|}{19.85}                         & \multicolumn{1}{c|}{25.36}                          & \multicolumn{1}{c|}{28.40}                          & \multicolumn{1}{c|}{39.26}                          & \multicolumn{1}{c|}{1.25}                         & \multicolumn{1}{c|}{1.62}                         & \multicolumn{1}{c|}{1.65}                         & 2.08                         \\ \midrule
\multicolumn{10}{c}{SDXL (SDXL + koala-700m)}                                                                                                                                                                                                                                                                                                                                                                                                                                                                                                  \\ \midrule
\multicolumn{1}{l|}{}                              & \multicolumn{1}{l|}{SDXL}                                               & \multicolumn{4}{c|}{159.35}                                                                                                                                                                                          & \multicolumn{4}{c}{7.45}                                                                                                                                                                 \\ \cmidrule(l){3-10} 
\multicolumn{1}{l|}{\multirow{-2}{*}{Non-Private}} & \multicolumn{1}{l|}{Hybrid SD}                                          & \multicolumn{2}{c}{31.87}                                                                                & \multicolumn{2}{c|}{63.74}                                                                                & \multicolumn{2}{c|}{1.45}                                                                             & \multicolumn{2}{c}{2.84}                                                         \\ \midrule
\multicolumn{1}{l|}{}                              & \multicolumn{1}{l|}{Vanilla OG}                                         & \multicolumn{1}{c|}{637.40}                        & \multicolumn{1}{c|}{956.10}                         & \multicolumn{1}{c|}{637.40}                         & \multicolumn{1}{c|}{956.10}                         & \multicolumn{1}{c|}{29.33}                        & \multicolumn{1}{c|}{43.16}                        & \multicolumn{1}{c|}{29.33}                        & 43.16                        \\
\multicolumn{1}{l|}{}                              & \multicolumn{1}{l|}{$\name$}                                            & \multicolumn{1}{c|}{127.48}                        & \multicolumn{1}{c|}{191.22}                         & \multicolumn{1}{c|}{254.96}                         & \multicolumn{1}{c|}{382.44}                         & \multicolumn{1}{c|}{5.92}                         & \multicolumn{1}{c|}{8.43}                         & \multicolumn{1}{c|}{11.35}                        & 16.90                        \\
\multicolumn{1}{l|}{}                              & \multicolumn{1}{l|}{$\enspace\textbf{+}$ cache}                         & \multicolumn{1}{c|}{97.62}                         & \multicolumn{1}{c|}{146.44}                         & \multicolumn{1}{c|}{180.42}                         & \multicolumn{1}{c|}{270.62}                         & \multicolumn{1}{c|}{4.08}                         & \multicolumn{1}{c|}{6.81}                         & \multicolumn{1}{c|}{7.98}                         & 11.28                        \\
\multicolumn{1}{l|}{}                              & \multicolumn{1}{l|}{\cellcolor[HTML]{9B9B9B}$\enspace\textbf{+}$ reuse} & \multicolumn{1}{c|}{\cellcolor[HTML]{9B9B9B}86.32} & \multicolumn{1}{c|}{\cellcolor[HTML]{9B9B9B}128.38} & \multicolumn{1}{c|}{\cellcolor[HTML]{9B9B9B}165.34} & \multicolumn{1}{c|}{\cellcolor[HTML]{9B9B9B}246.55} & \multicolumn{1}{c|}{\cellcolor[HTML]{9B9B9B}3.51} & \multicolumn{1}{c|}{\cellcolor[HTML]{9B9B9B}5.05} & \multicolumn{1}{c|}{\cellcolor[HTML]{9B9B9B}6.81} & \cellcolor[HTML]{9B9B9B}9.92 \\ \cmidrule(l){2-10} 
\multicolumn{1}{l|}{\multirow{-5}{*}{Private}}     & \multicolumn{1}{l|}{Total ($\enspace\textbf{+}$ device)}                & \multicolumn{1}{c|}{133.40}                        & \multicolumn{1}{c|}{175.46}                         & \multicolumn{1}{c|}{200.65}                         & \multicolumn{1}{c|}{281.86}                         & \multicolumn{1}{c|}{6.07}                         & \multicolumn{1}{c|}{7.61}                         & \multicolumn{1}{c|}{8.88}                         & 11.99                        \\ \bottomrule
\end{tabular}
}
    \label{tab:efficiency}
\end{table}

\subsection{Efficiency Improvement Breakdown}\label{append:efficiency-breakdown}

\begin{figure}[ht]
    \centering
    \includegraphics[width=0.8\linewidth]{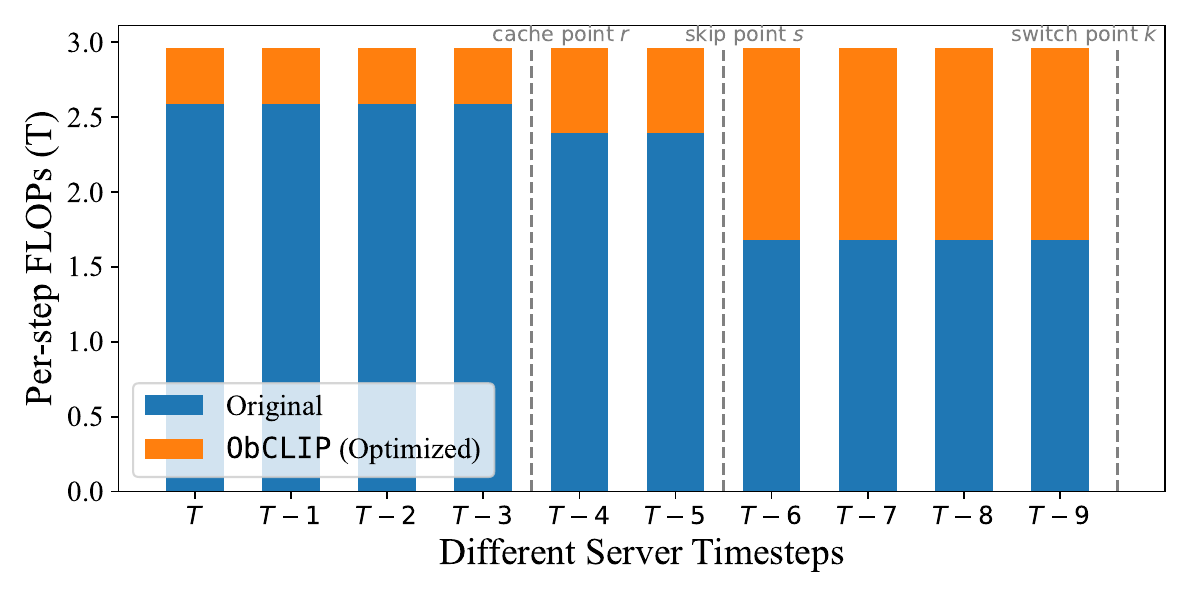}
    \caption{Per-step FLOPs (T) for SD-v1.4 with $k=10, r=4, s=6$.}
    \label{fig:flops-reduction}
\end{figure}

Take SD-v1.4 and BK-SDM-small hybrid generation with switch point $k=10$, cache point $r=4$, skip point $s=6$ as an example. We illustrate the per-step FLOPs of server-side denoising in Figure~\ref{fig:flops-reduction}, where the reduced computation FLOPs are highlighted in yellow. Prior to the cache point, only batch reuse of attention maps is enabled, resulting in a reduction of approximately 10\%. Subsequently, with the addition of temporal attention reuse and block skipping, reductions of about 20\% and nearly 50\% are achieved, demonstrating the effectiveness of the overall acceleration strategy.

\subsection{Visualization of Step-distilled Model Generation}\label{append:vis-lcm-sdxl}

\begin{figure}[ht]
    \centering
    \includegraphics[width=0.9\linewidth]{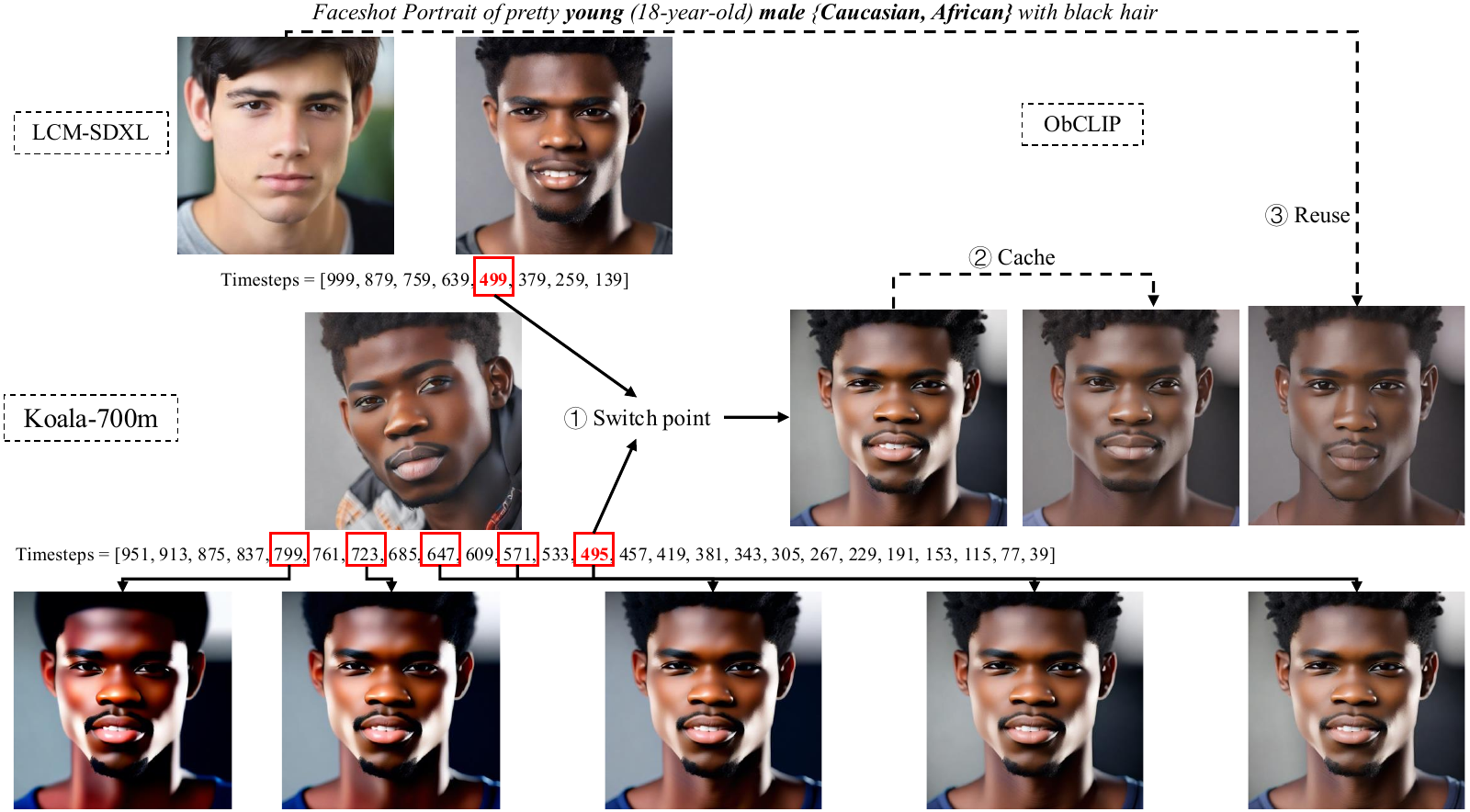}
    \caption{Visualization of LCM-SDXL + Koala-700m.}
    \label{fig:vis-lcm-sdxl}
\end{figure}

In this section, we present visualizations of images generated with the server-side LCM-SDXL (optimized using step distillation) and client-side Koala-700m models. Note that we use an 8-step LCM scheduler for LCM-SDXL and a 25-step DPM scheduler for Koala-700m.
Recall that as mentioned in Section~\ref{sec:quantitative}, we apply a timestep shift $\triangle t$ to align the timesteps between cloud and device as $t_{device} = t_{cloud} + \triangle t$. We here vary $\triangle t \in \{0, 2, 4, 6, 8\}$.
The input prompt is ``Faceshot Portrait of a pretty young (18-year-old) male \textit{African} with black hair.'' To demonstrate the effect of batch reuse, we choose a candidate prompt: ``Faceshot Portrait of a pretty young (18-year-old) male \textit{Caucasian} with black hair.''
As shown in Figure~\ref{fig:vis-lcm-sdxl}, the final generated images show poor fidelity when $\triangle < 6$, with $\triangle = 8$ yielding the best visual quality. Additionally, the text-image semantics are well aligned.

